\documentclass[twocolumn]{aastex631}
\shorttitle{Pitfalls of periodograms}
\shortauthors{H\"ubner et al.}

\usepackage{amssymb}
\usepackage{amsmath}
\usepackage{graphicx}
\usepackage{dcolumn}
\usepackage{xspace}

\usepackage{subfigure}
\let\tablenum\relax
\usepackage{siunitx}
\DeclareSIUnit \parsec {pc}

\usepackage[normalem]{ulem} %% for striking out text

\newcommand{\bilby}{\textsc{Bilby}\xspace}

\newcommand{\dynesty}{\textsc{dynesty}\xspace}

\newcommand{\Z}{\mathcal{Z}}
\newcommand{\M}{\mathcal{M}}
\renewcommand{\L}{\mathcal{L}}

\newcommand{\SPA}{School of Physics and Astronomy, Monash University, Clayton, VIC 3800, Australia}
\newcommand{\OzGravMonash}{OzGrav: The ARC Centre of Excellence for Gravitational Wave Discovery, Clayton, VIC 3800, Australia}
\newcommand{\SRON}{SRON Netherlands Institute for Space Research}

\newcommand{\CUA}{Physics Department, The Catholic University of America, Washington, DC, 20064, USA}
\newcommand{\SPL}{Solar Physics Laboratory, NASA Goddard Space Flight Center, Greenbelt, MD, 20771, USA}

\begin{document}

\title{Pitfalls of periodograms: The non-stationarity bias in the analysis of quasi-periodic oscillations}

\author{Moritz H\"ubner}
    \email{moritz.huebner@monash.edu}
    \affiliation{\SPA}
    \affiliation{\OzGravMonash}

\author{Daniela Huppenkothen}
    % \email{d.huppenkothen@sron.nl}
    \affiliation{\SRON}

\author{Paul D. Lasky}
    % \email{paul.lasky@monash.edu}
    \affiliation{\SPA}
    \affiliation{\OzGravMonash}

\author{Andrew R. Inglis}
    \affiliation{\CUA}
    \affiliation{\SPL}

% \linenumbers

\begin{abstract}

Quasi-periodic oscillations (QPOs) are an important key to understand the dynamic behavior of astrophysical objects during transient events like gamma-ray bursts, solar flares, and magnetar flares.
Searches for QPOs often use the periodogram of the time series and perform spectral density estimation using a Whittle likelihood function.
However, the Whittle likelihood is only valid if the time series is stationary since the frequency bins are otherwise not statistically independent.
We show that if time series are non-stationary, the significance of QPOs can be highly overestimated and estimates of the central frequencies and QPO widths can be overconstrained.
The effect occurs if the QPO is only present for a fraction of the time series and the noise level is varying throughout the time series.
This can occur for example if background noise from before or after the transient is included in the time series or if the low-frequency noise profile varies strongly over the time series.
We confirm the presence of this bias in previously reported results from solar flare data and show that significance can be highly overstated. 
Finally, we provide some suggestions that help identify if an analysis is affected by this bias.

\end{abstract}

\section{Introduction}\label{sec:introduction}
Quasi-periodic oscillations (QPOs)\footnote{In solar physics the term ``quasi-periodic pulsation'' (QPP) is preferred. We apply the term QPO generically throughout for simplicity.} are a common astrophysical phenomenon that are regularly observed across a variety of sources.
While there is ample discussion about how they emerge in their respective sources, it is worth re-examining existing techniques for their detection and characterization.
Observations of QPOs are scant for some objects such as in magnetar flares \citep{Strohmayer2004, Strohmayer2005, Israel2005, Watts2006, Huppenkothen2012, Huppenkothen2014, Huppenkothen2014a, Miller2019}, or contentious in others like gamma-ray bursts (GRBs) \citep{Cenko2010, DeLuca2010, Iwakiri2010, Morris2010, Tarnopolski2021}, and false detections may lead to unjustified theoretical inferences.
Moreover, even if there are ample detections of QPOs like in solar flares~\citep{Nakariakov2009, VanDoorsselaere2016, Zimovets2021} or X-ray binaries~\citep{Ingram2019}, making accurate inferences about their parameters and significance is essential for understanding the mechanisms that produce them.

There is an array of methods being applied to time series data to find QPOs.
The most common tests involve periodograms and are either based on outlier analyses or Bayesian tests~\citep{Vaughan2005, Vaughan2009}.
Other methods include wavelet analyses~\citep{Foster1996, Torrence1998, Lachowicz2010} and those based on Gaussian processes~\citep{Rasmussen2006, Foreman-Mackey2017, Covino2020, Zhu2020}.
% One major criticism of some of the methods used to analyze time series is the use of detrending ~\cite{Auchere2016}.
% Detrending is meant to remove low frequency components that arise due to the overall shape of the lightcurve of a transient, however it is likely to lead to spurious detections of QPOs because it can artificially enhance the power at certain frequencies.
% Due to these criticisms, detrending has fallen somewhat out of use for periodogram based analyses in favor of power law or broken-power law red noise models~\cite{Inglis2015, Inglis2016}, but is being used for wavelet analyses.

In this paper, we explore the effects that the non-stationary nature of a transient lightcurve, and the potential non-stationarity of a QPO within it, have on detection significance and characterization when methods are used that assume the underlying processes to be stationary. 
We show that we are likely to overestimate the significance of the quasi-periodic signal under these conditions.
This bias arises because neighboring bins in periodograms are only statistically independent for stationary time series.
We thus refer to this effect as the ``non-stationarity bias", which to the best of our knowledge has not been reported in the astrophysical literature to date.
We show that merely adding photon counting noise before or after a transient leads one to greatly overestimate the significance of a QPO and overconstrain the QPO parameters.
More critically, this effect also occurs if only the transient is selected for analysis but the QPO is only present for a fraction of it.

It is our intuition that astrophysicists would rather be conservative and include more of the time series that does not include the QPO, rather than ``cherry-picking'' the part of the lightcurve that appears most likely to contain it.
The erroneous reasoning behind this might be the belief that adding more noise to the signal should not increase the significance of the QPO.

We structure this paper as follows.
In Sec.~\ref{sec:c5:methods} we lay out the Bayesian methods and models that we use throughout.
Thereafter, in Sec.~\ref{sec:c5:pop}, we explain how the non-stationarity bias arises both on a conceptual level and with detailed mathematical arguments.
We show in Sec.~\ref{sec:c5:injection_study} based on simulated data that the non-stationarity bias exists empirically.
In Sec.~\ref{sec:c5:solar_flare} we show how the non-stationarity bias affects the analysis of solar flare lightcurves.
We conclude in Sec.~\ref{sec:c5:discussion} and provide some outlook on how alternative methods can potentially better handle non-stationary data sets.

\section{Methods}\label{sec:c5:methods}
In the following, we recapitulate the Fourier-based methods commonly used to analyze QPOs which are relevant to this study.
A general comprehensive overview on the topic of spectral density estimation for time series analysis is given in many popular textbooks such as chapter 7 in~\citet{Chatfield1976}, as well as in~\citet{VanderKlis1989, Barret2012} in the context of astrophysical lightcurves.
\subsection{Whittle likelihood}
A periodogram is an estimate of the power spectral density (PSD) of the signal based on a given time series $x(t)$.
We calculate the periodogram $I(f)$ as the absolute square of the discrete Fourier transform of the signal
\begin{equation}
    I(f) = \left|\sum_{i=1}^{N} x(t_i) \exp (-i2\pi ft_i) \right|^2 \, .
\end{equation}
Periodograms are established as a standard method in the search for QPOs in solar flares~\citep{Inglis2015, Inglis2016, Hayes2020} and QPOs in astrophysical transients elsewhere~\citep[e.g.~][]{Huppenkothen2012, Huppenkothen2014, Huppenkothen2014a, Huppenkothen2017, Miller2019}.
Since individual frequency bins $I(f_j) = I_j$ are calculated by taking the sum of the square of the normal-distributed real and imaginary parts of the Fourier series, it follows that they are $\chi^2_2$-distributed around the PSD $S(f_j) = S_j$~\citep{Whittle1951, Chatfield1976}.
This specific case of a $\chi^2_2$-distribution is identical to an exponential distribution around $S_j$
\begin{equation}\label{eq:exp_dist}
    p(I_j|S_j) = \frac{1}{S_j} \exp(-I_j/S_j) \, ,
\end{equation}
where $p(I_j|S_j)$ is the conditional probability to observe the power $I_j$ given an underlying PSD $S_j$.
This relation is generically true for any individual frequency bin in any periodogram.
Assuming that all bins are statistically independent from another, we obtain the Whittle likelihood function by taking the product over all $N/2$ frequency bins of a periodogram corresponding to a time series with $N$ points
\begin{equation}\label{eq:whittle}
    L(I|S) =  \prod_{j=1}^{N/2}\frac{1}{S_j} \exp(-I_j/S_j) \, .
\end{equation}
We emphasize here that this is only true in the stationary limit, and this will lead to biased estimates of the PSD in general, as we show in Sec.~\ref{sec:c5:pop}.
Nevertheless, Eq.~\ref{eq:whittle} is the standard likelihood that is used for spectral density estimation.

\subsection{Models of power spectra}
Many astrophysical transients show excess power at low frequencies, and it is often assumed that this can be modeled using a red-noise process (see e.g.~\citealt{Huppenkothen2012,Inglis2015, Inglis2016, Miller2019, Ingram2019, Broomhall2019}).
One basic noise model that is commonly used is a combination of a red-noise power law with amplitude $A$ and spectral index $\alpha$, and a white noise amplitude $C$~\citep{Inglis2015, Inglis2016},  i.e.
\begin{align}\label{eq:model_s0}
    S^{\mathrm{RW}}(f) &= S^{\mathrm{R}}(f) + S^{\mathrm{W}} \\ 
    &= A f^{-\alpha} + C \, ,\nonumber
\end{align}
where we use the $\mathrm{R}$ and $\mathrm{W}$ superscripts as short-hand for ``red noise'' and ``white noise'' respectively. 
This spectral shape usually emerges as a combination from the overall structure of the transient, and additional variability on smaller time scales.
The shape is thus not due to a stationary process but rather due to a combination of unknown deterministic and non-stationary stochastic processes adding up to mimic a red noise spectrum~\citep{Huppenkothen2012}.

A common way to model a QPO is to add a Gaussian or Lorentzian enhancement to $S^{\mathrm{RW}}$, e.g.
\begin{align}\label{eq:model_s1}
    S^{\mathrm{RWQ}}(f) &= S^{\mathrm{RW}}(f) + S^{\mathrm{Q}}(f)\\
    &= S^{\mathrm{RW}}(f) + \frac{B}{\pi\sigma} \frac{\sigma^2}{(f - f_0)^2 + \sigma^2} \, ,\nonumber
\end{align}
where $B$ is the QPO amplitude, $f_0$ is its mean frequency, and $\sigma$ is the half-width half-maximum scale parameter.
Explicitly modeling the red noise component is important as the QPO is likely to overlap with the red noise dominated part of the PSD.
Not considering the red noise dominated part of the PSD may thus lead to false positives.

There are extensions to the noise model, such as the broken or bent power law (B superscript), that can fit more structured red noise and are often a better fit to the data
\begin{equation}\label{eq:model_s2}
    S^{\mathrm{BW}}(f) = A f^{-\alpha_1}  \left(1 + \left(\frac{f}{\delta} \right)^{\alpha_2 - \alpha_1}\right) + C \, ,
\end{equation}
where $\alpha_{1,2}$ are the power law indices present before and after the break frequency $\delta$ where the power law changes, and we also enforce $\alpha_2 < \alpha_1$ to avoid degeneracies.
There are also various different formulations of bent or broken power laws with various degrees of smoothness.

\subsection{Model selection and parameter estimation}
There are several ways to assert the significance of a QPO when using spectral density estimation.
A widely used frequentist way to detect QPOs is to use outlier statistics.
As a first step, one fits the PSD using a model that does not contain a QPO, e.g. Eq.~\ref{eq:model_s0} or Eq.~\ref{eq:model_s2}.
Thereafter, every frequency bin is checked if its amplitude exceeds a set probability threshold based on the $\chi^2_2$-distribution in Eq.~\ref{eq:exp_dist}.
One also has to account for the number of trials, i.e. the number of frequency bins tested by applying a Bonferroni-correction~\citep{Bonferroni1936}.
Additionally, it is possible to rebin the periodogram into a smaller number of bins.
For QPOs with widths larger than the width of a single frequency bin, rebinning may be useful because it increases the significance of the QPO relative to the noise continuum.

Alternative to the outlier statistics, one can cast this as a model Bayesian selection problem where we find the preferred model to fit our data, in our case Eqs.~\ref{eq:model_s0},~\ref{eq:model_s1}, and~\ref{eq:model_s2}.
There are multiple ways to quantify model preference in Bayesian statistics.
One approach that has been used in solar physics \citep{Ireland2015, Inglis2015, Inglis2016} is to calculate the Bayesian Information Criterion ($BIC$)
\begin{equation}\label{eq:BIC}
    BIC = k\ln(n) - 2\ln (L_{\max})
\end{equation}
where $k$ is the number of free parameters (e.g. three for $S^{\mathrm{RW}}$, six for $S^{\mathrm{RWQ}}$, and five for $S^{\mathrm{BW}}$), $n=N/2$ is the number of data points, and $L_{\max}$ is the maximum likelihood value.
This method is relatively cheap computationally as the calculation of the maximum likelihood can be obtained with relatively few likelihood evaluations using a maximization algorithm.
A lower $BIC$ indicates a better fit to the data, thus the difference in $BIC$ for $S^{\mathrm{RW}}$ and $S^{\mathrm{RWQ}}$ is a measure of statistical significance of the QPO,
\begin{align}\label{eq:delta_bic}
    \Delta BIC &= BIC(S^{\mathrm{RWQ}}) - BIC(S^{\mathrm{RW}})\\ 
    &= (k_{\mathrm{RWQ}} - k_{\mathrm{RW}})\ln(N/2) - 2\ln \frac{L_{\max}(I|S^{\mathrm{RWQ}})}{L_{\max}(I|S^{\mathrm{RW}})} \,  ,\nonumber
\end{align}
where $\Delta BIC < 0$ would indicate that $S^{\mathrm{RWQ}}$ is preferred and vice versa.
We can also perform model selection via calculation of Bayes factors $BF$.
To understand Bayes factors, we start with Bayes' theorem
\begin{equation}
    p(\theta|d, S) = \frac{\pi(\theta|S)L(d|\theta, S)}{Z(d| S)} \, ,
\end{equation}
where $\theta$ are the model parameters, $d$ are the data, i.e. the periodogram in our case, $p$ is the posterior probability of the parameters, $\pi$ is the prior probability of the parameters, $L$ is the likelihood of the data given the parameters, and $Z$ is the evidence, or fully marginalized likelihood.
All these probabilities are conditioned on a model we want to evaluate, e.g. a PSD $S$.
The Bayes factor comparing two models is the ratio of their evidences.
For example, the Bayes factor comparing $S^{\mathrm{RWQ}}$ and $S^{\mathrm{RW}}$ is 
\begin{equation}
    BF = \frac{Z(d| S^{\mathrm{RWQ}})}{Z(d| S^{\mathrm{RW}})}\, .
\end{equation}
It thereby measures the relative odds of the underlying data to have been produced by either model, though it does not measure if the model itself is a good fit to the data, similar to the $BIC$.
The evidence is calculated by rearranging and integrating Bayes' theorem
\begin{equation}\label{eq:evidence}
    Z(d| S) = \int \pi(\theta)L(d|\theta, S) d\theta \, .
\end{equation}
Evaluating this integral is much more computationally challenging than calculating the $BIC$, but can be achieved thanks to improvements in algorithms such as nested sampling~\citep{Skilling2006}, and accessible software implementations such as \bilby~\citep{Speagle2019, Ashton2018, Romero-Shaw2020}.
The Bayes factor obtained via evidence calculation is seen as the superior standard for model selection in Bayesian statistics since it involves prior beliefs about the distribution of parameters~\citep{Weakliem1999}.
The penalty factor $k\ln(n)$ for the $BIC$ also increases with the size of the data set and can thus favor overly simplistic models~\citep{Weakliem1999, Gelman2013}.
Since the $BIC$ relies on a point estimate of the likelihood, it is also a less reliable measure if the likelihood has multiple modes.
Additionally, nested sampling provides posterior distributions on the model parameters which are interesting in their own right.

Bayes factors and $BIC$s work best if the underlying models are truly discrete, but $S^{\mathrm{RW}}$ is a special case of both $S^{\mathrm{RWQ}}$ and $S^{\mathrm{BW}}$.
There are reasonable criticisms on the use of Bayes factors or the $BIC$ for cases like this as $S^{\mathrm{RW}}$ is a subset of both $S^{\mathrm{RWQ}}$ and $S^{\mathrm{BW}}$, and thus not a truly discrete model~\citep{Gelman2013}.
In this case, the Bayes factor depends on prior choices of the parameters specific to $S^{\mathrm{RWQ}}$ and $S^{\mathrm{BW}}$.
Alternatively, we could use tests that measure if the posterior of the QPO amplitude $B$ is inconsistent with being zero and provides a better fit to the data.
Nevertheless, we use Bayes factors as prior choices since they will yield a constant difference in our Bayes factors and will not distort the overall trend when we investigate the non-stationarity bias.
We also consider the effects on the $\Delta BIC$s where it is instructive.

For all the Bayes factor calculations and posterior samples in this publication we use \dynesty via the \bilby interface~\citep{Ashton2018,Speagle2019, Romero-Shaw2020}.

\section{Pitfalls of Periodograms}\label{sec:c5:pop}

\begin{figure}
\centering
\subfigure{\includegraphics[width=\linewidth]{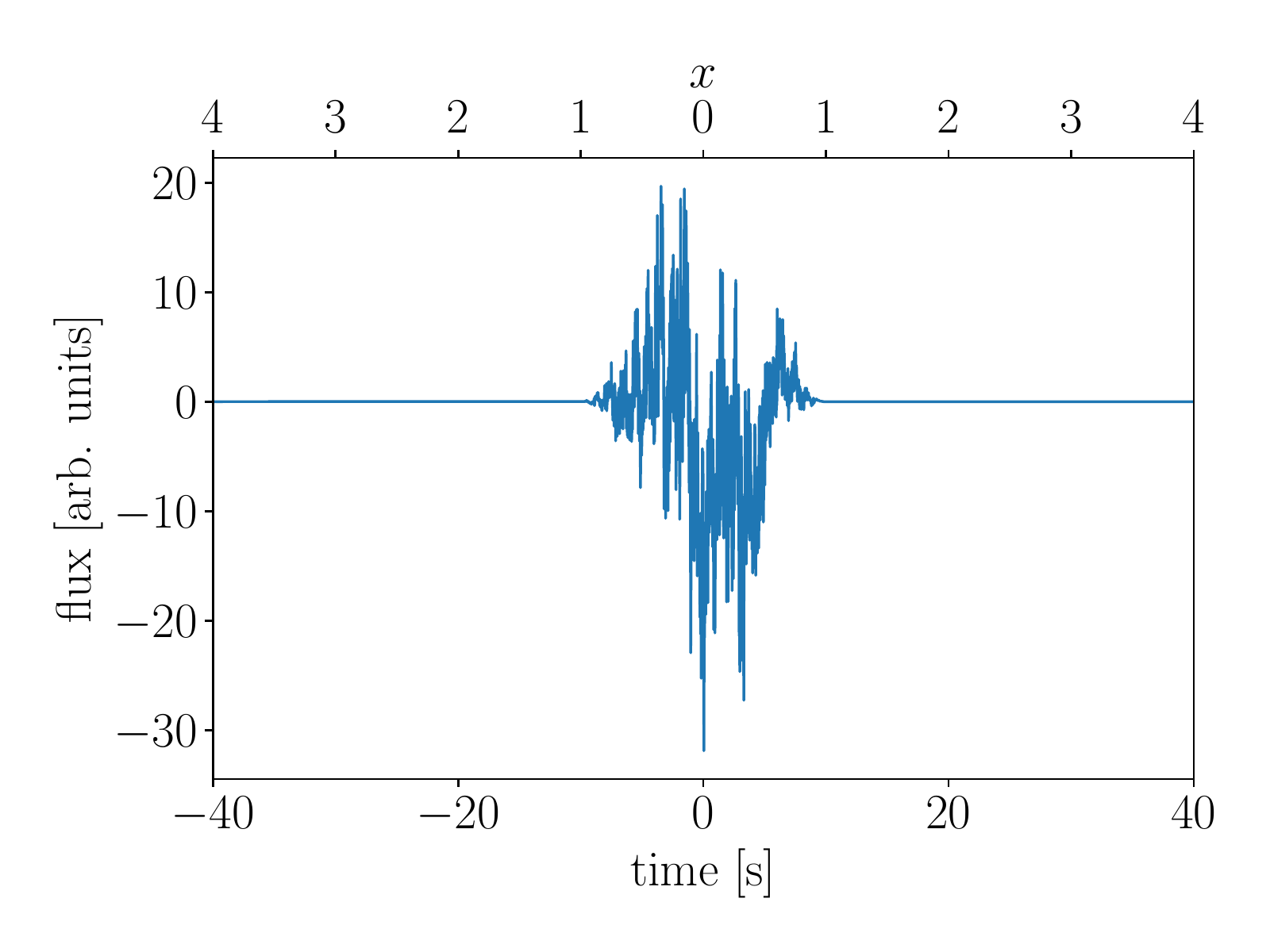}}
\subfigure{\includegraphics[width=\linewidth]{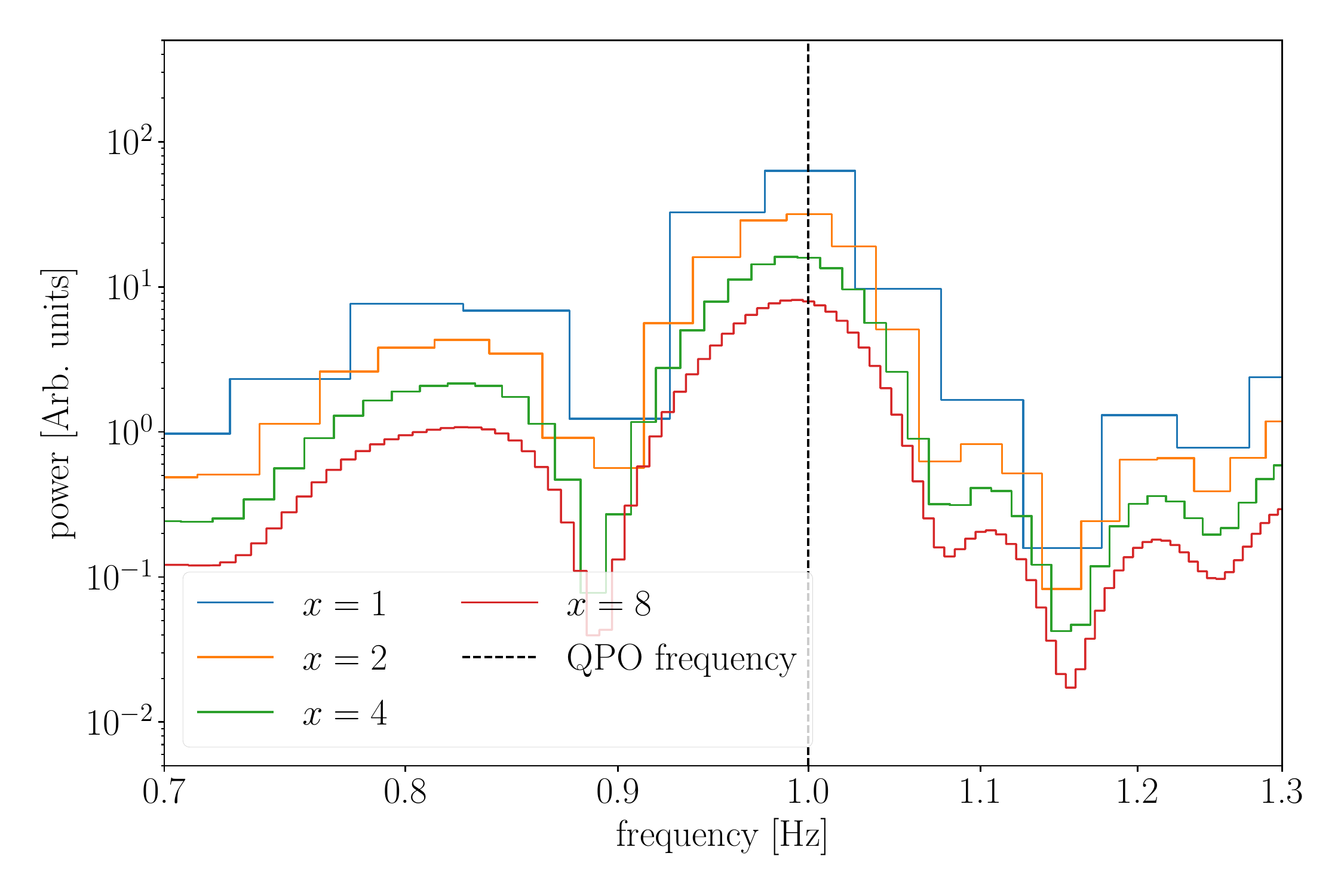}}
\caption{Effects of extending a time series with zeros.
    In the top panel, we show a stationary time series between $\SI{-10}{\second}$ and $\SI{10}{\second}$ that we multiply with a Hann window to ensure a smooth turn on from zero. 
    We extended this time series with zeros on either end up to a total duration of \SI{200}{\second}.
    The upper x-axis shows how the extension factor $x$ maps onto the time axis.
    Taking only the stationary data from \SI{-10}{\second} to \SI{10}{\second} corresponds to $x=1$, i.e. no extension has been applied.
    The extension factor in this case implies that we select data from $-10x$\,\si{\second} to $10x$\,\si{\second}.
    In the bottom panel, we show the periodogram given different extension factors, with the blue curve representing the stationary data ($x=1$).
    The other colors (orange, green, red) show the effect of extending the time series with zeros on either end.
    In these curves, neighboring bins are not statistically independent, which manifests itself in the emerging smooth structure of the periodogram.
    The bins are also decreasing inversely proportional to $x$ since we do not apply a normalization.
}
\label{fig:zero_padding_effects}
\end{figure}

Assume a discrete time series of length $T$ with $N$ data points containing a QPO and noise.  
The underlying stochastic process is characterized in the frequency domain as a PSD $S(f)$, which can be estimated by calculating a periodogram $I(f)$.
We recall that the value in any specific of the $N/2$ frequency bins in the periodogram is proportional to the absolute square of the Fourier amplitude.
% The periodogram is not a ``consistent'' estimator of the PSD, which means that the uncertainty on $I(f)$ would not decrease if we the time series was longer.
% Instead, obtaining a longer time series by e.g. observing the same process for a longer time, decreases the spacing between adjacent frequency bins.
% For spectral density estimation this means that we need to fit the periodogram with model of $S(f)$ to get a better estimate.

In the top panel of Fig.~\ref{fig:zero_padding_effects} we extend a stationary time series by appending zeros on either end until we reach a length $xT$ with $xN$ data points, where we call $x$ the \textit{extension factor}.
As we display in the bottom panel, this increases the number of frequency bins in the periodogram to $xN/2$, while decreasing the periodogram to $I(f)/x$ if we do not apply a normalization.
The non-stationarity bias occurs here because an increased number of frequency bins boosts the calculated significance of a QPO in terms of the $\Delta BIC$ or $\ln BF$ between, e.g. $S^{\mathrm{RWQ}}$ and $S^{\mathrm{RW}}$.

The example displayed in Fig.~\ref{fig:zero_padding_effects} is not realistic in practice but serves to illustrate the effect.
A more realistic scenario occurs if the stationary time series is instead extended with low amplitude white noise and the QPO is within the red noise dominated part of the PSD.
This raises the white noise level relative to red noise and the QPO.
Yet it still increases the significance of the QPO, which is determined by its amplitude relative to the dominant red noise level.
Such a scenario might occur if we overselect data surrounding a transient

Another scenario is that the overall time series is non-stationary, and instead, the QPO and the overall noise processes vary, or switch on and off, during a selected time window.
For example, the QPOs reported in the 2004 SGR1806-20 hyperflare are directly associated with specific rotational phases of the magnetar~\citep{Strohmayer2005, Israel2005, Strohmayer2006, Watts2006, Huppenkothen2014a, Miller2019}.
Parts of the time series might only contribute low levels of noise and thereby increase the number of bins without suppressing the QPO in the periodogram.
These parts, therefore, increase the significance of the QPO without having delivered any actual information about it.

We now turn towards a more mathematical explanation of the non-stationarity bias.
Let us consider how we can calculate the SNR of a QPO when red and white noise are present\footnote{The argument here holds for arbitrary noise.}.
Given a single frequency bin $f_j$, this is 
\begin{equation}\label{eq:snr}
    \rho(f_j) = \frac{S^{\mathrm{Q}}(f_j)}{S^{\mathrm{RW}}(f_j)} \, ,
\end{equation}
where we denote the SNR as $\rho$.
Assuming all frequency bins in a periodogram are statistically independent, the SNR adds in quadrature
\begin{equation}\label{eq:snrtot}
    \rho_{\mathrm{tot}} = \sqrt{\sum_{j=1}^{N/2} \rho(f_j)^2} \, ,
\end{equation}
where we have summed over the $N/2$ bins in the periodogram and $\rho_{\mathrm{tot}}$ denotes the total SNR of the QPO.

\subsection{QPOs in stationary noise}\label{sec:qpos_in_stationary_noise}
As the most simple scenario, we consider a stationary time series with $N$ elements made up of a QPO and some arbitrary noise. 
We denote the SNR of this time series as $\rho^{N}_{\mathrm{tot}}$.
Suppose we look at an extended version of this time series with $xN$ elements, which we create with the same QPO and noise process.
We intuitively expect the SNR to increase proportionally to $\sqrt{x}$ and indeed, we calculate
\begin{align}\label{eq:snr_stationary_extension}
\begin{split}
    \rho^{xN}_{\mathrm{tot}} &= \sqrt{\sum_{j=1}^{xN/2} \rho^{x}(f_j)^2} \\
                             &= \sqrt{x\sum_{j=1}^{N/2} \rho(f_j)^2} \\
                             &= \sqrt{x} \rho^{N}_{\mathrm{tot}}\, , 
\end{split}
\end{align}
where we write the SNR at a given frequency in the extended time series as $\rho^{x}(f_j)$, which is the same as in the original time series.
We also use the $N$ and $xN$ superscripts for the SNRs to indicate whether we are looking at the original or extended time series. 
Note that we have implicitly changed how the $f_j$ are indexed between the steps.
Additionally, we assume in the first step that the PSD is only slowly varying between frequency bins and we can therefore approximate the newly added frequency bins by their closest neighbors from the original PSD.

Alternatively, let us consider what happens when we consider a non-stationary extended time series in which the noise is present over $xN$ elements but the QPO is only present for $N$ elements. 
In that case, $S^{\mathrm{RW}}$ is the same as in the original time series whereas $S^{\mathrm{Q}}$ gets reduced to $S^{\mathrm{Q}}/x$ as we are effectively ``diluting'' the QPO, similarly to how the periodograms decrease in amplitude with increasing $x$ in the lower panel of Fig.~\ref{fig:zero_padding_effects}.
We calculate Eq.~\ref{eq:snrtot} for the extended time series again
\begin{align}
\begin{split}
    \rho^{xN}_{\mathrm{tot}}
    &= \sqrt{\sum_{j=1}^{xN/2} \rho^{x}(f_j)^2} \\
    &= \sqrt{x\sum_{j=1}^{N/2} \rho^{x}(f_j)^2} \\
    &= \sqrt{x\sum_{j=1}^{N/2} \rho(f_j)^2 / x^2} \\
    &= \frac{1}{\sqrt{x}} \rho^{N}_{\mathrm{tot}}\, ,
\end{split}
\end{align}
where we used the first same step as in Eq.~\ref{eq:snr_stationary_extension} and used the fact that the QPO is suppressed by the extension in the second step.
These first two scenarios are fairly intuitive, analyzing more of the same stationary process should naturally boost the significance, whereas only adding more noise will dilute the QPO and reduce the significance.

Let us now consider what happens if we extend the time series with zeros on either end, which is illustrated in Fig.~\ref{fig:zero_padding_effects}.
This scenario is not truly physical, as we always expect some background noise.
However, it is an instructive example to understand the effects of non-stationarities.
Extending with zeros implies both $S^{\mathrm{RWQ}} \rightarrow S^{\mathrm{RWQ}}/x$ and $S^{\mathrm{RW}} \rightarrow S^{\mathrm{RW}}/x$, so $\rho^{x}(f) = \rho(f)$ and thus following the same steps as before
\begin{equation}
    \rho^{xN}_{\mathrm{tot}} = \sqrt{x} \rho^{N}_{\mathrm{tot}} \, , 
\end{equation}
which is identical to Eq.~\ref{eq:snr_stationary_extension}.
This is a curious result as extending a time series by zeros clearly should not increase the SNR nor the significance of a signal.
The resolution to this seeming paradox is that it is invalid to assume that SNR adds in quadrature as we did in Eq.~\ref{eq:snr_stationary_extension}.
By extending the time series with zeros, we added bins to the periodogram that are not statistically independent.

A further consequence is that the product in the Whittle likelihood becomes invalid.
Due to the extended segment length both the PSD and the periodogram need to be divided by a factor of $x$.
Therefore, we can rewrite the likelihood as
\begin{align}
    L^{xN}(I|S(f|\theta)) &= \prod_{j=1}^{xN/2}\frac{x}{S_j(\theta)} \exp(-I_j/S_j(\theta)) \\
                  &\approx x^{xN/2} \left( \prod_{j=1}^{N/2}\frac{1}{S_j(\theta)} \exp(-I_j/S_j(\theta)) \right)^x \, ,\nonumber
\end{align}
where $\theta$ is the set of parameters in the PSD and we use the $N$ and $xN$ superscipts on the likelihoods to indicate if we are looking at the original or extended time series.
Again, we assume that the PSD is slowly varying with frequency and thus statistically non-independent bins are reasonably approximated by the closest frequency bins obtained from the stationary time series.
The factor $x^{xN/2}$ that we have introduced does not matter since it is a constant.
Hence we can write
\begin{equation}\label{eq:extended_log_l}
    \ln L^{xN}(I|S(\theta)) \propto x \ln L^N(I|S(\theta))\, .
\end{equation}
This means $\ln L^{xN}(I|S(\theta)) \approx x \ln L^N(I|S(\theta))$, i.e. we have steepened the log likelihood function by a factor of $x$
\begin{equation}
    \frac{\partial \ln L^{xN}(I|S(\theta))}{\partial \theta} \approx x\frac{\partial \ln L^N(I|S(\theta))}{\partial \theta} \, .
\end{equation}
This steepening means that inferred posterior distributions are generally tighter which leads to overconstrained parameter estimates.
We calculate how this changes the Bayesian information criterion
\begin{align}
\begin{split}
    \Delta BIC^{xN} &= (k_{\mathrm{RWQ}} - k_{\mathrm{RW}}) \ln(xN/2) - \\ 
    &2x\ln \frac{L^N_{\max}(I|S^{\mathrm{Q}})}{L^N_{\max}(I|S^{\mathrm{RW}})} \\ 
    &= x \Delta BIC + \\ 
    &\, (k_{\mathrm{RWQ}} - k_{\mathrm{RW}}) (\ln x - (x-1) \ln (N/2))\, ,
\end{split}
\end{align}
where $k_{\mathrm{RWQ}}=6$ and $k_{\mathrm{RW}}=3$ are the number of free parameters for the respective models, and $xN$ is used as a superscript.
Thus, $\Delta BIC$ is approximately proportional to $x$ or $N$ since the impact of the $\ln x$ and constant terms are minimal.
% given sufficiently large $x$.
% For $|x - 1| \ll 1$ we have $\ln x \approx 1 + x$ and thus $\Delta BIC^{xN} \approx x \left( \Delta BIC + (k_{\mathrm{RWQ}} - k_{\mathrm{RW}}) \right) + (k_{\mathrm{RWQ}} - k_{\mathrm{RW}})$.
The impact this has on the Bayes factor is not straight-forward to calculate in the general case.
Since likelihood ratios and $\Delta BIC$s can be understood as a related quantity of the $\ln BF$, we expect a similar approximately linear behavior, which we find empirically in Sec.~\ref{sec:c5:injection_study}.

\subsection{QPOs in non-stationary red noise}
Assume a stationary time series with red noise, white noise, and a QPO that sits in the red-noise dominated part of the PSD.
If we extend this time series with zeros, the argument in Sec.~\ref{sec:qpos_in_stationary_noise} holds and the SNR calculated with Eq.~\ref{eq:snrtot} grows proportional to $\sqrt{x}$.
Extending with white noise instead of zeros will also lead to the SNR growing with $\sqrt{x}$ as long as the QPO remains in the red noise dominated part of the PSD.
To show this we define $f_{\mathrm{break}}$ as the breaking frequency where white noise and red noise are of equal magnitude
\begin{equation}
    f_{\mathrm{break}} =\left( \frac{C}{A} \right)^{-1/\alpha} \, ,
\end{equation}
which means that for $f \ll f_{\mathrm{break}}$ red noise dominates whereas white noise dominates for $f \gg f_{\mathrm{break}}$.
If our QPO has $f_0 < f_{\mathrm{break}}$ it mainly competes with red noise in terms of SNR, thus a rising level of white noise will not meaningfully reduce the SNR until $f_0 \approx f_{\mathrm{break}}$.
We can write this as
\begin{equation}
    \rho_{\mathrm{tot}} \approx \rho_{\mathrm{tot}}(f < f_{\mathrm{break}}) \,,
\end{equation}
where $\rho_{\mathrm{tot}}(f < f_{\mathrm{break}})$ is the total SNR just based on frequencies less than the breaking frequency.
If we extend the time series with white noise or zeros by a factor of $x$, we suppress the red noise term by $1/x$ and shift the breaking frequency
\begin{equation}\label{eq:break_x}
    f_{\mathrm{break}}^x = \left( \frac{Cx}{A} \right)^{-1/\alpha} = x^{-1/\alpha}f_{\mathrm{break}} \, ,
\end{equation}
where the superscript $x$ indicates that we are looking at the extended time series.
We can equate Eq.~\ref{eq:break_x} to $f_0$ and re-arrange to calculate the extension factor $x_{\mathrm{break}}$ for which the QPO will be in the white-noise dominated part of the periodogram
\begin{equation}\label{eq:x_break}
    x_{\mathrm{break}} = \left(\frac{f_{\mathrm{break}}}{f_0}\right)^{\alpha} = \frac{A}{C}f_0^{-\alpha} \, .
\end{equation}
Therefore, we are prone to artificially increase the SNR by extending our time series with white noise up to an extension factor of $x_{\mathrm{break}}$.
Conversely, if our QPO mainly competes with white noise $f_0 > f_{\mathrm{break}}$, we use the argument in Sec.~\ref{sec:qpos_in_stationary_noise} to show that we suppress the SNR with increasing $x$
\begin{align}
\begin{split}
    \rho_{\mathrm{tot}, x} &= \sqrt{\sum_{j=1}^{xN/2}\left(\frac{S^{\mathrm{Q}}(f_j)/x}{S^{\mathrm{W}}} \right)^2}\\ 
    &\approx \frac{1}{\sqrt{x}}\sqrt{\sum_{j=1}^{N/2}\left(\frac{S^{\mathrm{Q}}(f_j)}{S^{\mathrm{W}}} \right)^2} \\
    &\approx \frac{\rho_{\mathrm{tot}}}{\sqrt{x}}\, ,
\end{split}
\end{align}
where we have used the fact that the PSD is slowly varying with frequency in the first approximation, and that the frequencies where the QPO exceeds the noise continuum are all white-noise dominated in the second approximation.
% In this case the artificial increase SNR by $\sqrt{x}$ due to the increase in the number of frequency bins by a factor of $x$ is overwhelmed by the decrease in $S^{\mathrm{Q}}$ and thus SNR by a factor of $1/\sqrt{x}$.

The most obvious case where this could become a practical problem is when we have to decide where to start and end the segment.
Not being aware of this bias may lead us to overselect the data by $\mathcal{O}(10\%)$ and infer a somewhat higher SNR and significance, which by itself is not a big issue.
But the issue becomes much worse in practice if the QPO itself is non-stationary and only appears for a part of the transient, which implies that selecting the entire transient may already be overextending the segment in which the QPO is present.

In principle, extending the time series with red noise instead of white noise should not cause the same issue as we would expect the same suppression by $1/\sqrt{x}$ that we have for white noise in Sec.~\ref{sec:qpos_in_stationary_noise}.
However, in real transients, the low-frequency noise continuum arises not due to a stationary noise process but due to a combination of deterministic and non-stationary stochastic processes.
Different parts of the time series might add up roughly to a single power law or a broken power law, but smaller segments within the transient may have vastly different shapes.
For example, the rising and falling edge of the transient may create different power laws, we may have segments that are fairly flat and thus mostly add white noise, and so on.
Additionally, deterministic aspects of the time series, if not properly subtracted, are not appropriately modeled with a Whittle likelihood as we show in Sec.~\ref{sec:c5:non_stat_qpo_in_transient}.
The deterministic parts of the lightcurve can not be safely removed by filtering or smoothing methods as they are prone to create an artificial structure that looks like oscillatory behavior~\citep{Auchere2016}.
Clearly, it is not possible to write down a rigorous treatment of this rather large class of possible constellations in which non-stationarity bias could occur.

We note here that using the outlier analysis is not prone to false detections due to the non-stationarity bias since we are testing neighboring bins individually instead of combining SNR from neighboring bins.
The opposite is the case, since we add more bins by having extended the time series, it is harder to exceed the Bonferroni-corrected significance level.
However, in practice periodograms are sometimes rebinned to add up neighboring bins \citep[e.g.~][]{Huppenkothen2014}.
The motivation behind this is to combine the SNR from neighboring bins, making it easier to detect QPOs.
Thus, rebinned outlier analyses are prone to the same non-stationarity bias that affects the Bayesian methods as well.

\section{Simulated data}\label{sec:c5:injection_study}
To demonstrate the impact of non-stationarities empirically, we analyze simulated data using a variety of setups.
% We consider three specific scenarios.
% First, we consider a combination of white noise and a QPO, that we extend with either more white noise or zeros to make it non-stationary.
% Second, we create a transient using a deterministic flare shape that mimics a red-noise continuum in the PSD, and add stationary white noise and a non-stationary QPO that is present for a fraction of the transient.
% Third, we create a stationary time series using red noise, white noise, and a QPO, which we then extend with either more red noise, white noise or zeros.
For all studies on simulated data, we assume white noise levels to be constant along with the time series.
We discuss the impact of the Poissonian nature of the photon counting process in Appendix~\ref{sec:c5:det_process_poiss}.

\subsection{Setup}\label{sec:c5:setup}
To produce the time domain data, we use the algorithm by~\citet{Timmer1995}, as it is implemented in the \bilby software package.
Concretely, this works by creating a white noise frequency series by randomly drawing both amplitudes and phases from a normal distribution, coloring the noise by multiplying it with the square root of the PSD and then applying an inverse Fourier transform to obtain a stationary time series. 
In general, we create a continuous, non-stationary time series by using a combination of addition, concatenation, and convolution with window functions.
For example, we start by creating a stationary time series using the QPO model and red noise $S^{\mathrm{RQ}} = S^{\mathrm{R}} + S^{\mathrm{Q}}$.
We apply a Hann window to ensure a smooth turn on from zero.
This helps us to avoid Fourier artifacts such as side-lobes in the periodogram which can appear if discontinuities are present.
Next, we create a much longer time series containing just white noise using the ~\citet{Timmer1995} method, and add the $S^{\mathrm{RQ}}$ time series in the center.
We are also interested in what happens when we extend the time series with zeros instead of white noise.
To do this we mask the parts containing just white noise and set those values to zero.
In all cases, we center the part of the time series in which the QPO is present.
This is to prevent the Hann window, which is applied when calculating the periodogram, from diminishing the amplitude of the red noise and QPO components.
Note that this is separate from the Hann window we use to ensure the smooth turn-on from zero.

For all simulated data in this section, we assume a constant white noise level throughout the time series and place the QPO in the center of the time series between \SI{-10}{\second} and \SI{10}{\second}.
We use a sampling frequency of \SI{40}{\hertz} and extend the time series up to \SI{200}{\second} ($x = 10$) or \SI{400}{\second} ($x = 20$) depending on the scenario.
The specific parameters for each simulation are listed in the tables in App.~\ref{sec:c5:injection_param_table}.
We discuss the effect of non-stationary white noise in App.~\ref{sec:c5:det_process_poiss}.
All parameters and priors are listed in Tabs.~\ref{tab:priors} and~\ref{tab:injection_params}.

\subsection{Non-stationary QPO in white noise}\label{sec:c5:non_stat_qpo_in_wn}
We start by considering a combination of a QPO and white noise where the setup is effectively identical to what we describe in~\ref{sec:c5:setup}, but we use $S^{\mathrm{Q}}$ instead of $S^{\mathrm{RQ}}$ to omit red noise.
We consider both scenarios discussed in Sec.~\ref{sec:qpos_in_stationary_noise} where we either extend the time series with zeros or with white noise.
In terms of the parameter inference, we deviate from the standard practice of using $S^{\mathrm{RWQ}}(f)$ as there is no red noise component.
Instead we compare $S^{\mathrm{WQ}}$ against the white noise hypothesis $S^{\mathrm{W}} = C$.

\begin{figure}
\centering
\subfigure{\includegraphics[width=\linewidth]{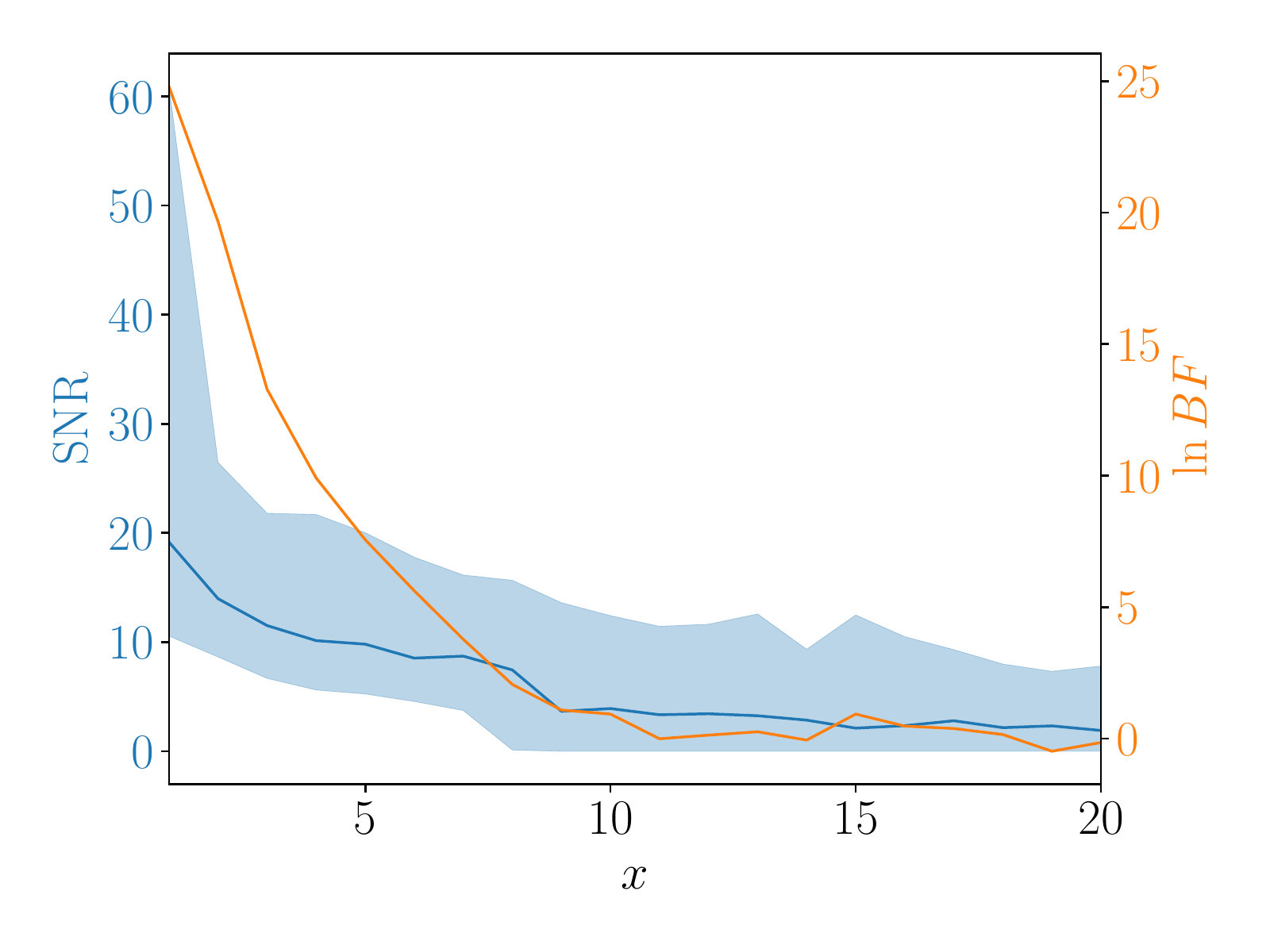}}
\subfigure{\includegraphics[width=\linewidth]{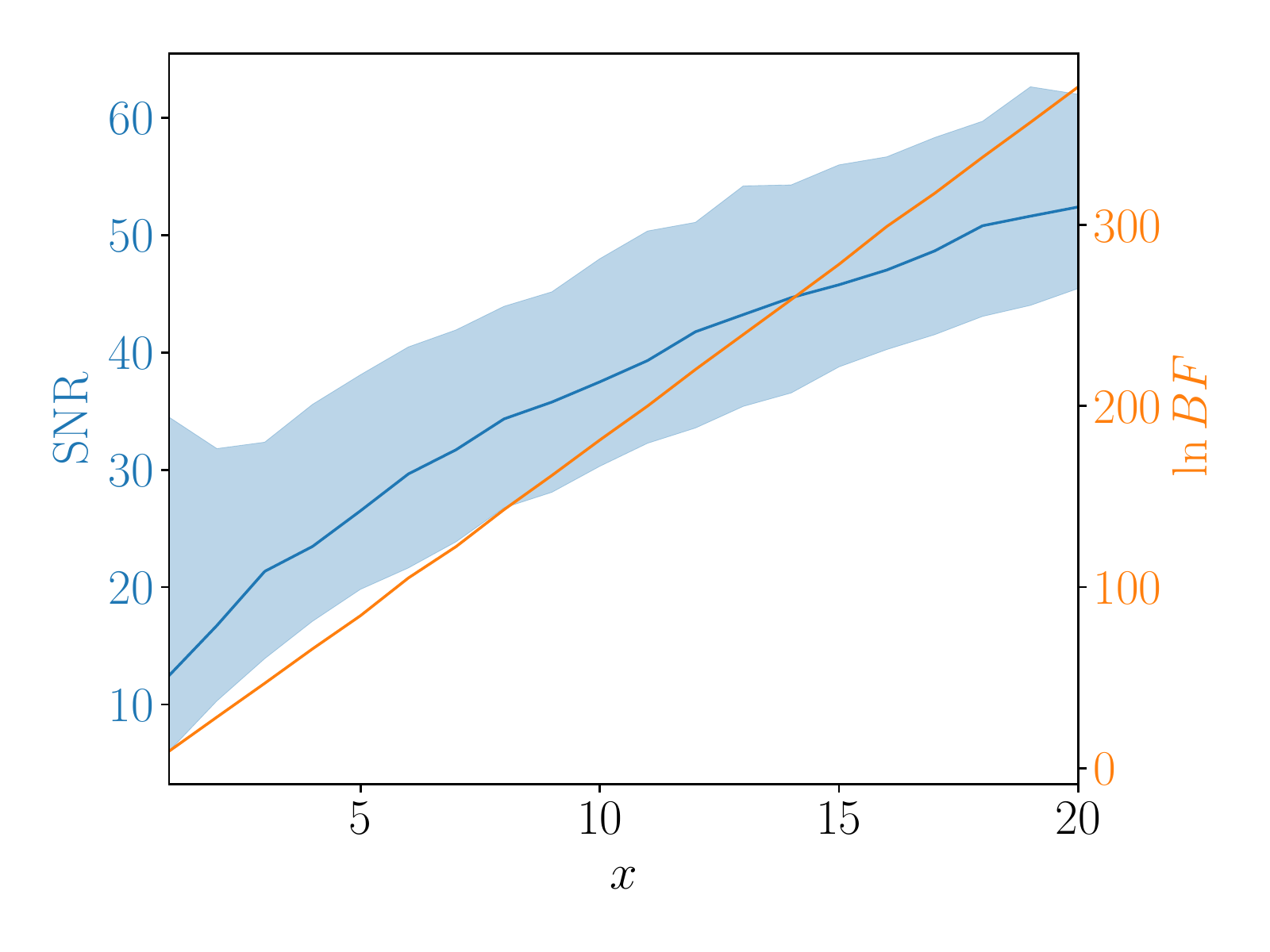}}

\caption{SNR (blue) and $\ln BF$ (orange) vs the extension factor for a simulated QPO plus white noise signal that is extended with white noise (top) or zeros (bottom).
The solid blue line displays the SNR of the maximum likelihood point and the shaded blue region displays the SNR's 90\% credible interval.
In the top panel both SNR and $\ln BF$ decrease with increasing $x$ which corresponds to the QPO vanishing in noise.
In the bottom panel the $\ln BF$ increases almost perfectly linearly due to the increased number of frequency bins that make up the QPO, whereas the SNR increases roughly with the square root.
}
\label{fig:non_stat_qpo_wn_ln_bf_snr}
\end{figure}

First, we consider what happens when we extend the time series with white noise.
In principle, the results should follow our discussion in Sec.~\ref{sec:qpos_in_stationary_noise} for QPOs in the white-noise-dominated part of the periodogram, i.e. the SNR should decrease with $1/\sqrt{x}$ and the $\ln BF$ should decrease roughly inversely to $x$.
The results, which we display in the top panel of Fig.~\ref{fig:non_stat_qpo_wn_ln_bf_snr}, are in agreement with our expectations.

We now consider the case where we extend the time series with zeros, similar to what we did to produce Fig.~\ref{fig:zero_padding_effects}.
As we discussed in Sec.~\ref{sec:c5:pop}, we expect the SNR to increase with $\sqrt{x}$ and the $\ln BF$ roughly linearly.
As we show in the bottom panel of Fig.~\ref{fig:non_stat_qpo_wn_ln_bf_snr}, this is almost perfectly the case for the $\ln BF$ and is also qualitatively true for the SNR.

\subsection{Non-stationary QPO in a simple transient}\label{sec:c5:non_stat_qpo_in_transient}
\begin{figure}
\centering
\subfigure{\includegraphics[width=\linewidth]{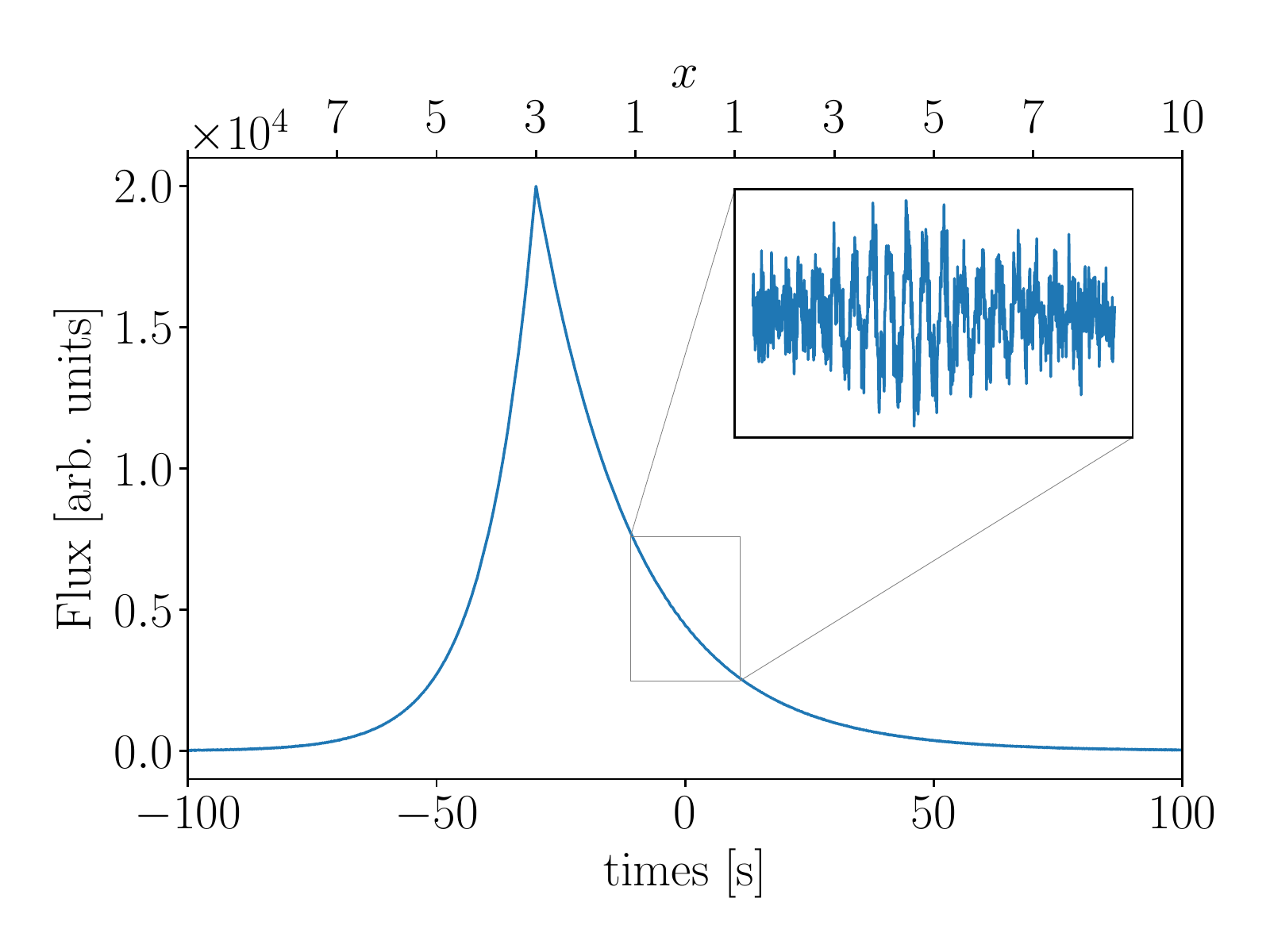}}
\subfigure{\includegraphics[width=\linewidth]{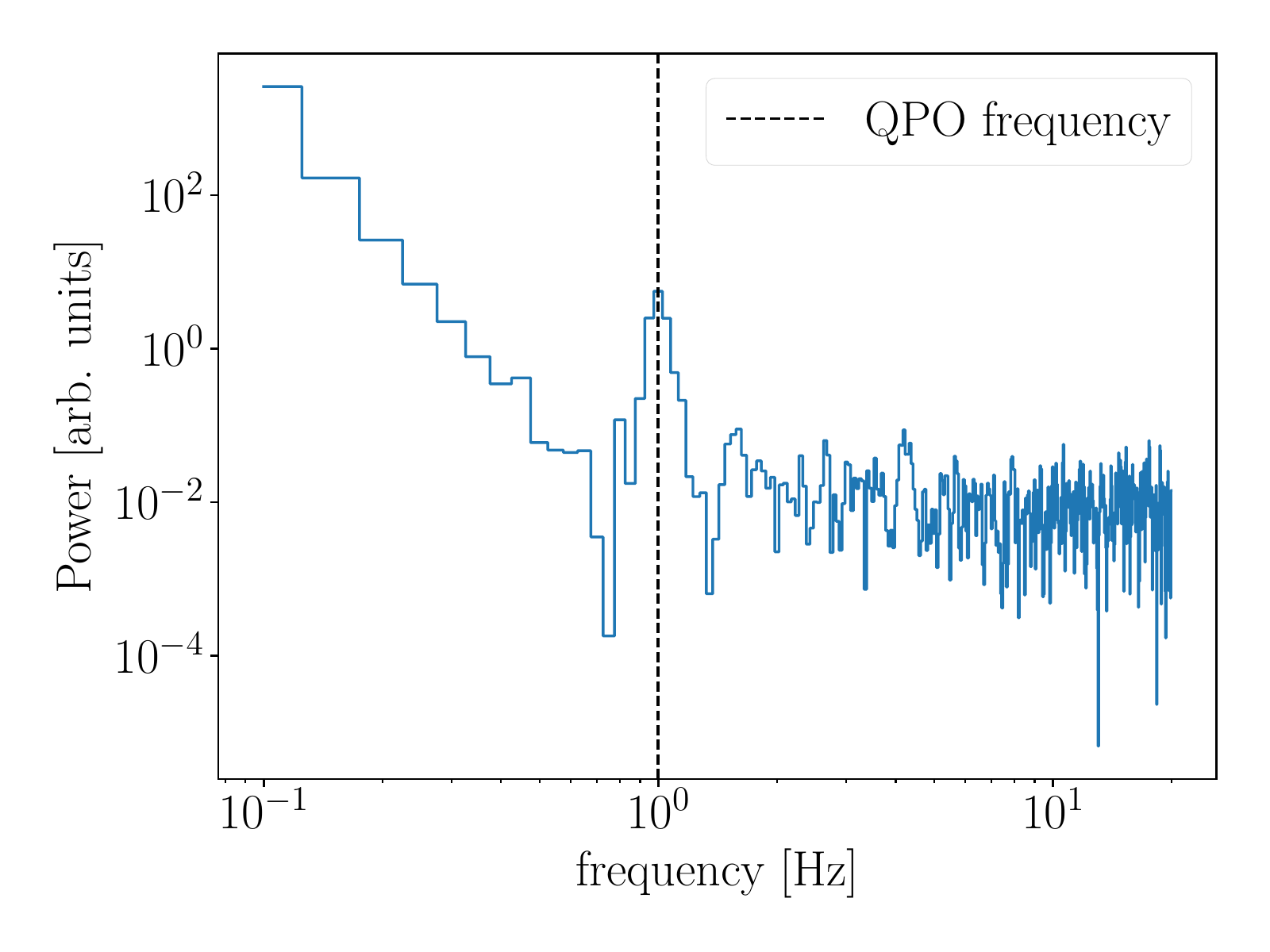}}

\caption{Simulated data of a non-stationary QPO in a deterministic transient flare shape in the presence of white noise. 
The time series (top) shows the dominant transient and the QPO that is present for \SI{20}{\second} on the tail of the flare (inset).
We remove the deterministic trend of the flare in the inset to make the QPO easily visible. 
The periodogram (bottom) shows a visible QPO at \SI{1}{\hertz} and displays a low-frequency continuum that arises due to the transient flare shape.
The periodogram corresponds only to the segment of the lightcurve that contains the QPO.
Note that there is much less variability than we would expect from the $\chi^2_2$-distribution for the bins at low frequencies since the noise continuum is not due to a stationary stochastic process but due to a deterministic process.
}
\label{fig:inj_qpo_in_transient_data}
\end{figure}

We investigate the effect of analyzing a non-stationary QPO that appears on top of a deterministic transient flare shape in the presence of white noise.
The time series is displayed in the top panel of Fig.~\ref{fig:inj_qpo_in_transient_data}.
The flare shape mimics a red-noise continuum that overlaps with the QPO in the periodogram (bottom panel of Fig.~\ref{fig:inj_qpo_in_transient_data}), though it is not well described by a Whittle likelihood in the periodogram since it is not a stochastic process.
We see that there is conspicuously little fluctuation at low frequencies in Fig.~\ref{fig:inj_qpo_in_transient_data}.
For the flare shape we use a continuous exponential rise and fall described by
\begin{align}
    f(t) = A_{\mathrm{flare}}
    \begin{cases} 
    \exp \left( \frac{t - t_0}{\tau_{\mathrm{r}}} \right) \, &\mathrm{if} \,  t < t_0\\
    \exp \left( -\frac{t - t_0}{\tau_{\mathrm{f}}} \right) \, &\mathrm{if} \, t \geq t_0
    \end{cases}
\end{align}
where $A_{\mathrm{flare}}$ is the amplitude, $t_0$ is the peak time, and $\tau_{\mathrm{r}}$ and $\tau_{\mathrm{f}}$ are the rise and fall timescales, respectively.
This model is similar to the FRED model which is popular to fit time series of gamma-ray bursts~\citep{Norris2005}.
The advantage of using an exponential shape for this simulation is that it produces a power law in the periodogram which matches our red noise model.
Since the flare is deterministic, low frequencies in the periodogram are not well described by a Whittle likelihood.
To create this time series, we repeat the steps in Sec.~\ref{sec:c5:non_stat_qpo_in_wn}, though we start with different parameters listed in Tab.~\ref{tab:injection_params}, and add the deterministic flare shape in the last step.

\begin{figure}
    \centering
    \includegraphics[width=\linewidth]{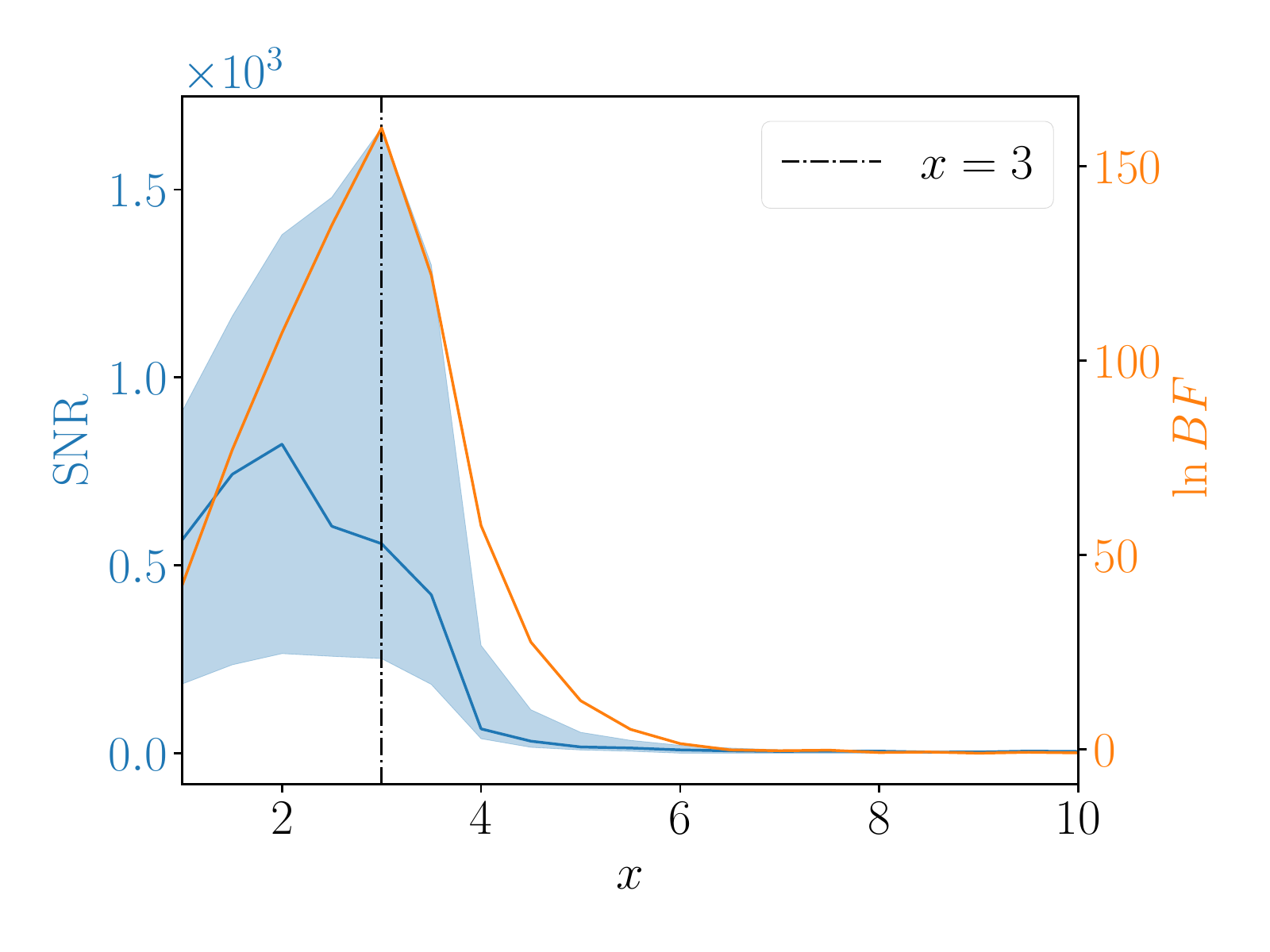}
    \caption{
    SNR (blue) and $\ln BF$ (orange) vs the extension factor for the QPO shown in Fig.~\ref{fig:inj_qpo_in_transient_data}.
    The solid blue line displays the SNR of the maximum likelihood point and the shaded blue region displays the SNR's 90\% credible interval.
    Both the $\ln BF$ and the upper edge of the SNR's credible interval peak at $x=3$ (black dashed vertical line), which corresponds to the point where the selected data segment exceeds past the peak of the flare.
    The strong differences in $\ln BF$ and SNR across extension factors shows that differences data selection can greatly influence the results.}
    \label{fig:inj_qpo_in_transient_ln_bfs}
\end{figure}

We perform the same test as in the previous section of extending the time series starting from the \SI{20}{\second} segment in which the QPO is present (inset in the top panel of Fig.~\ref{fig:inj_qpo_in_transient_data}).
It is less intuitive why non-stationarity bias may occur and how it will manifest in this scenario. 
Extending beyond the peak of the flare might reduce the apparent significance as it will provide the strongest contribution to the low-frequency noise continuum.
At the same time, we extend towards the tail of the flare, which contributes little more than white noise.
As we can see in Fig.~\ref{fig:inj_qpo_in_transient_ln_bfs}, the SNR and $\ln BF$ do indeed increase sharply until $x=3$, which corresponds to the point where we extend the selected window to the peak of the transient.
Past this point low-frequency contributions from the rising edge of the transient obscure the QPO in the periodogram and quickly lead to the QPO becoming undetectable for $x > 7$.
We note that using a different set of parameters to create the data can easily create situations where the $\ln BF$ continues to increase past $x=3$.

\subsection{Non-stationary QPO in non-stationary red noise}\label{sec:c5:non_stat_qpo_in_non_stat_red_noise}

\begin{figure}
\centering
\subfigure{\includegraphics[width=\linewidth]{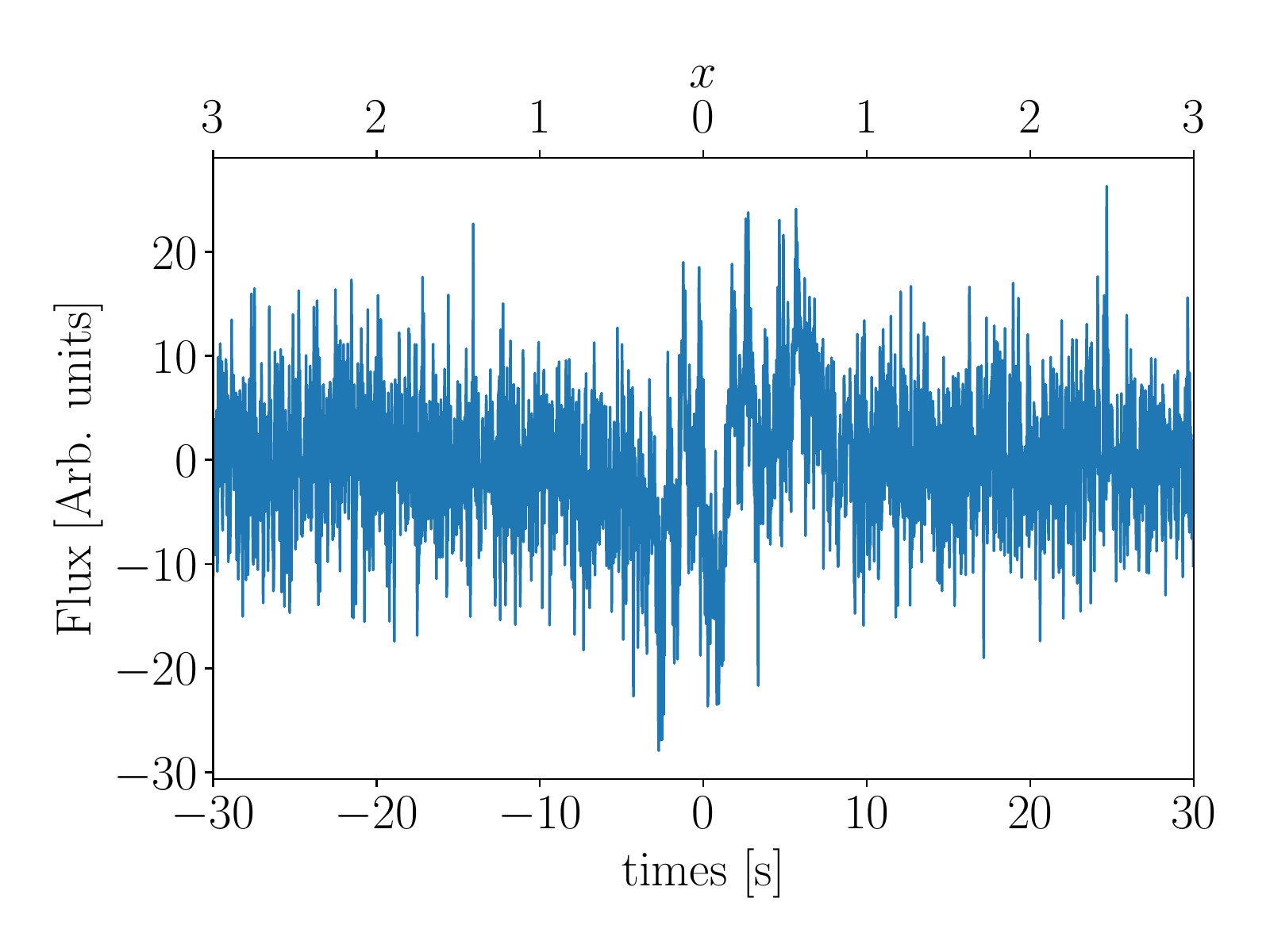}}
\subfigure{\includegraphics[width=\linewidth]{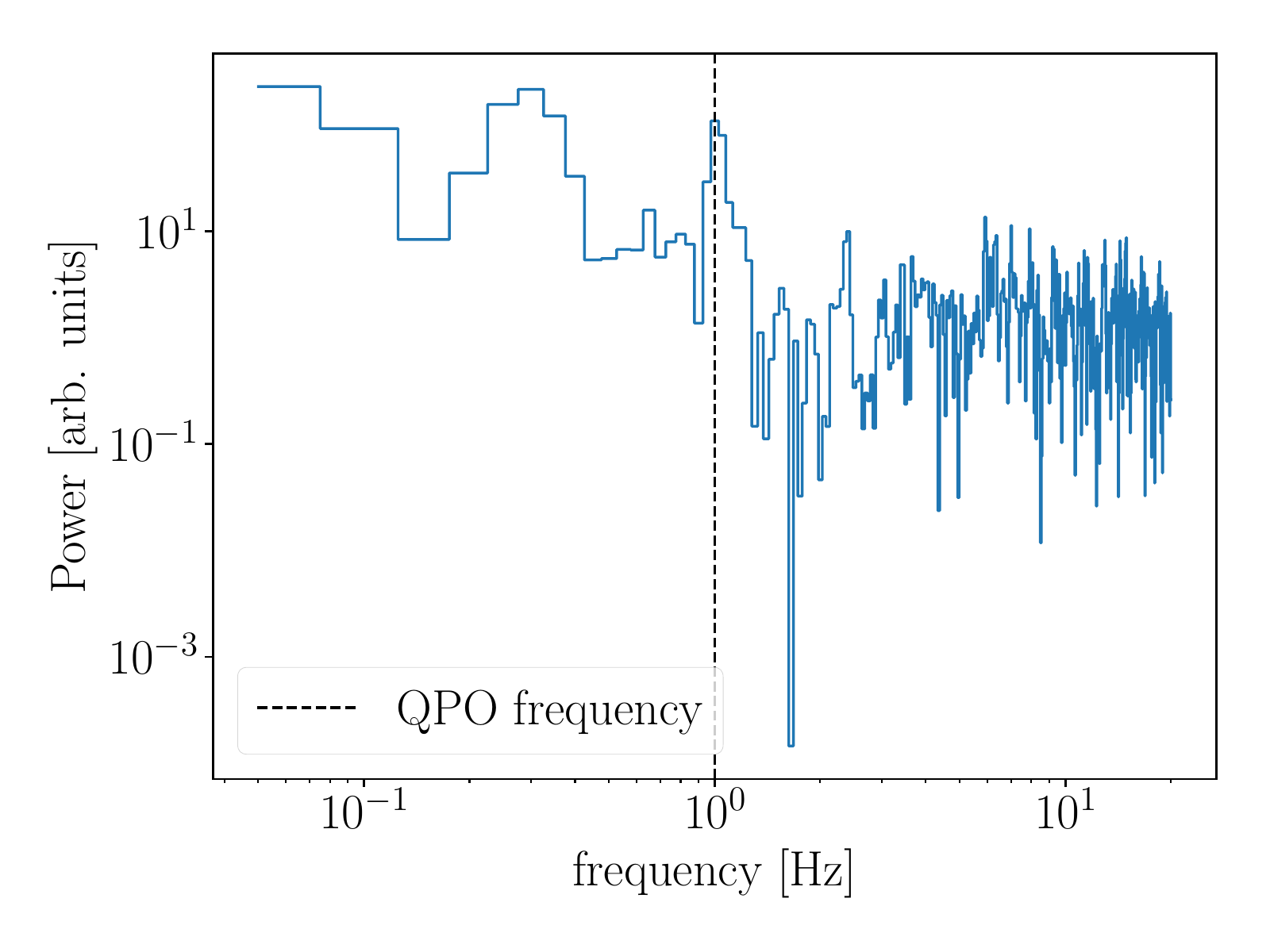}}

\caption{Simulated data of a non-stationary QPO and red noise in surrounding stationary white noise. 
In the time series (top) we show how QPO and red noise are transitioning smoothly into the surrounding part of the time series that is pure white noise ($|t| > \SI{10}{\second}$) due to the Hann window that we have applied.
In the periodogram (bottom) for $|t| \leq \SI{10}{\second}$ the QPO is clearly visible by eye.
}
\label{fig:inj_rn_qpo_in_white_noise}
\end{figure}

In this scenario, we create a time series using a combination of red noise, white noise, and a QPO using the $S^{\mathrm{RWQ}}$ model.
We extend this time series with either white noise or zeros.
Fig.~\ref{fig:inj_rn_qpo_in_white_noise} displays the time series for $|t| < \SI{30}{\second}$ (top panel) and the periodogram as obtained for $|t| < \SI{10}{\second}$.

\begin{figure}[ht]
\centering
\subfigure{\includegraphics[width=\linewidth]{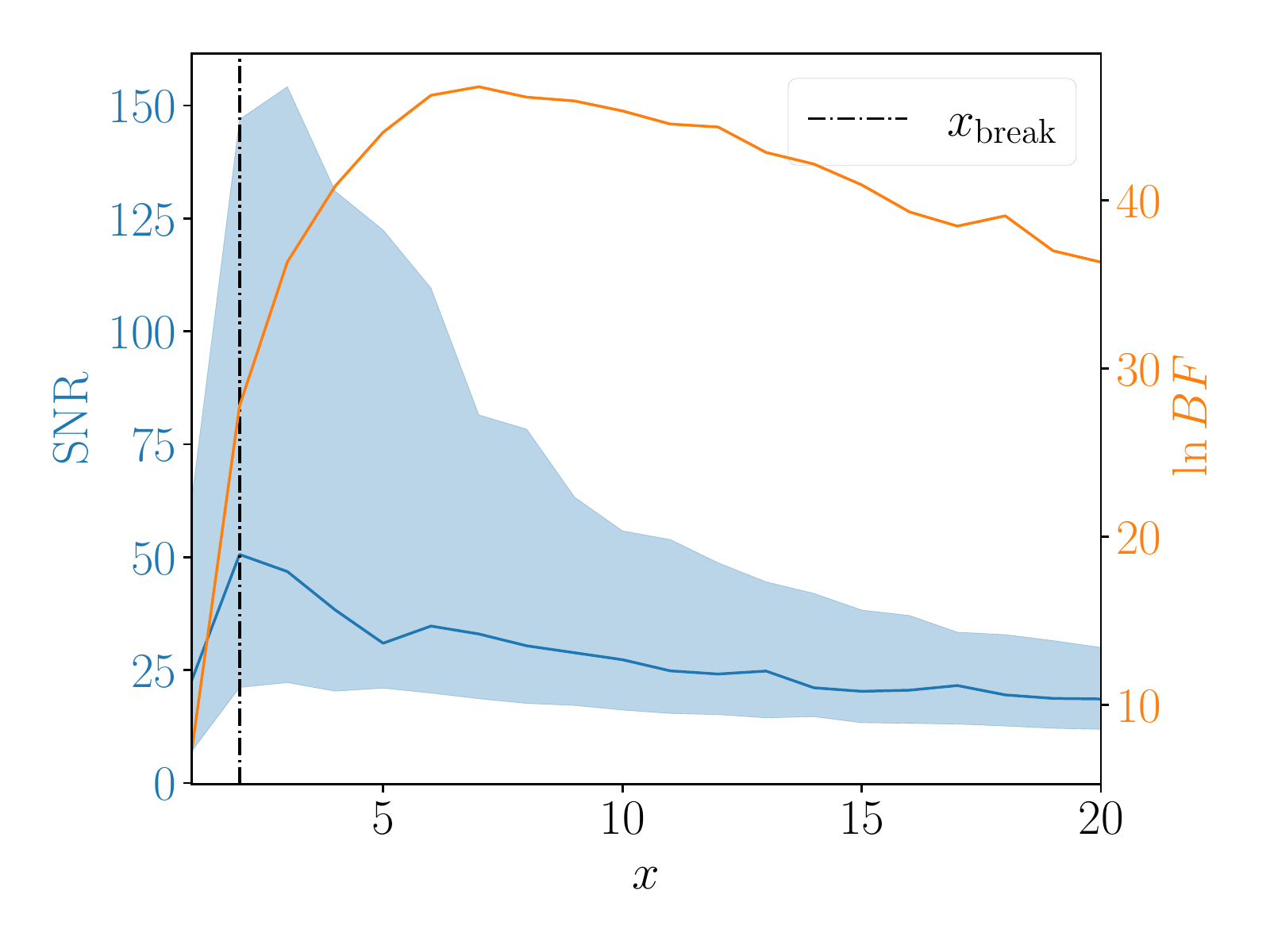}}
\subfigure{\includegraphics[width=\linewidth]{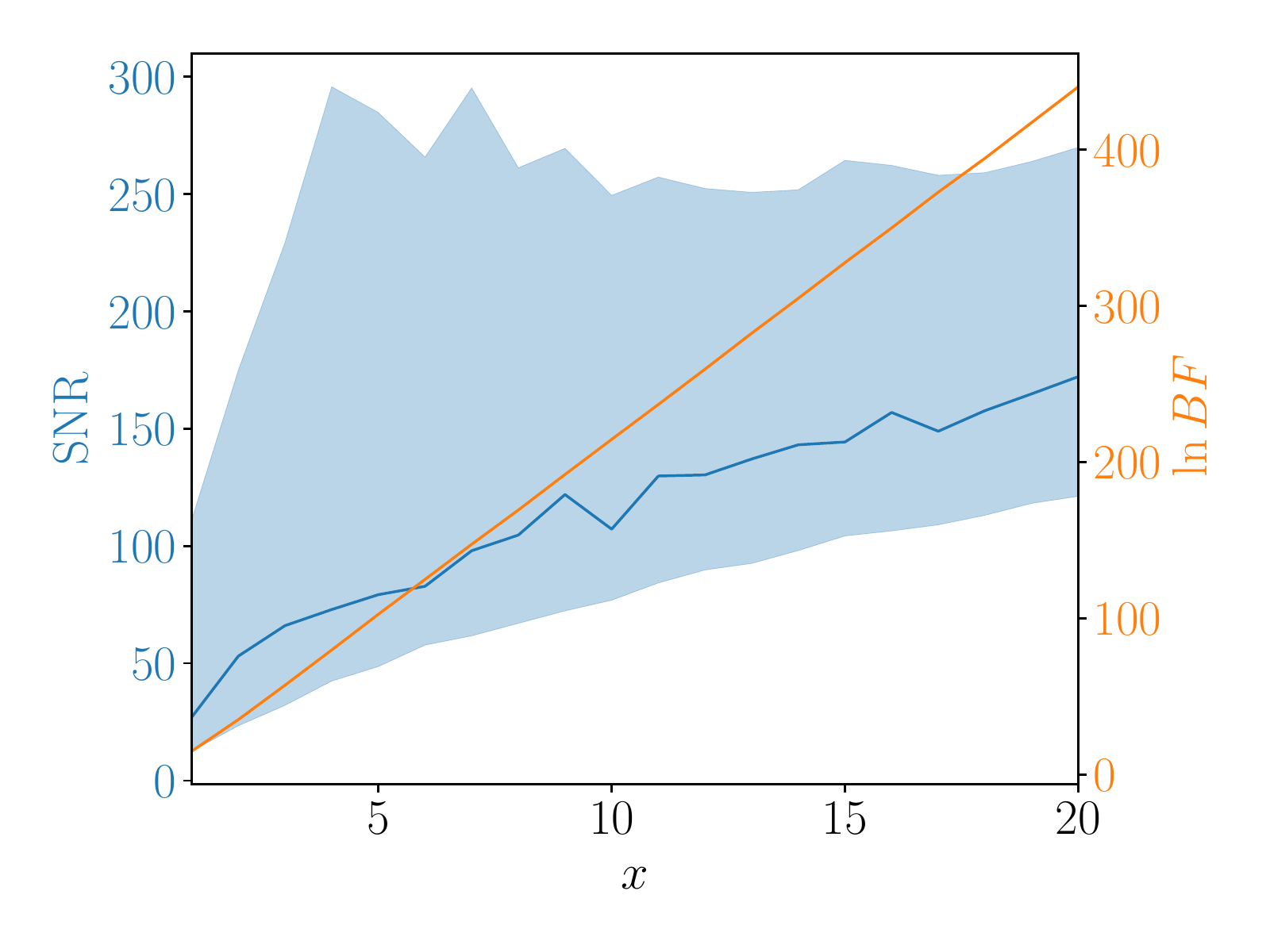}}
\caption{SNR (blue) and $\ln BF$ (orange) vs the extension factor for simulated data of a non-stationary QPO and red noise with white noise as described in Sec.~\ref{sec:c5:non_stat_qpo_in_non_stat_red_noise} for extension with white noise (top) and zeros (bottom).
The solid blue line displays the SNR of the maximum likelihood point and the shaded blue region displays the SNR's 90\% credible interval.
}
\label{fig:snr_vs_extension}
\end{figure}

In the case of extending with zeros, we expect the same relation as in Sec.~\ref{sec:c5:non_stat_qpo_in_wn}, which we recover in the bottom panel of Fig.~\ref{fig:snr_vs_extension}.
On the other hand, when we extend with white noise the SNR first rises quickly until the extension factor reaches around $x_{\mathrm{break}} = 2$ (see Eq.\ref{eq:x_break}) and then turns towards a slow descent proportional to $1/\sqrt{x}$.
We note that the behavior in the case of $x < x_{\mathrm{break}}$ for white noise is not well captured, but greater extension factors would be less realistic and the simulated data set aims to qualitatively show all effects.

We use models $S^{\mathrm{RW}}$ and $S^{\mathrm{RWQ}}$ as described earlier to calculate posterior distributions, Bayes factors, and $\Delta BIC$s.
As we show in the bottom panel of Fig.~\ref{fig:snr_vs_extension}, the $\ln BF$ increases almost perfectly linearly when we extend with zeros.
In the top panel, we see that if we extend with white noise the $\ln BF$ also increases with $x$ even past $x_{\mathrm{break}}$, though turns around eventually.
The $\ln BF$ may be increasing past $x_{\mathrm{break}}$ if the QPO feature in the periodogram is better fit by the model for higher $x$ compared to the feature's shape at $x_{\mathrm{break}}$.

We have shown that overall we can see a non-stationarity bias if white noise remains stationary, but red noise vanishes for part of the lightcurve.
This is in principle different from the effect in Sec.~\ref{sec:c5:non_stat_qpo_in_transient}, as the red noise is due to a stochastic rather than a deterministic process.

\subsection{Tests of stationarity}\label{sec:c5:chi_square}
\begin{figure}
\centering
\subfigure{\includegraphics[width=\linewidth]{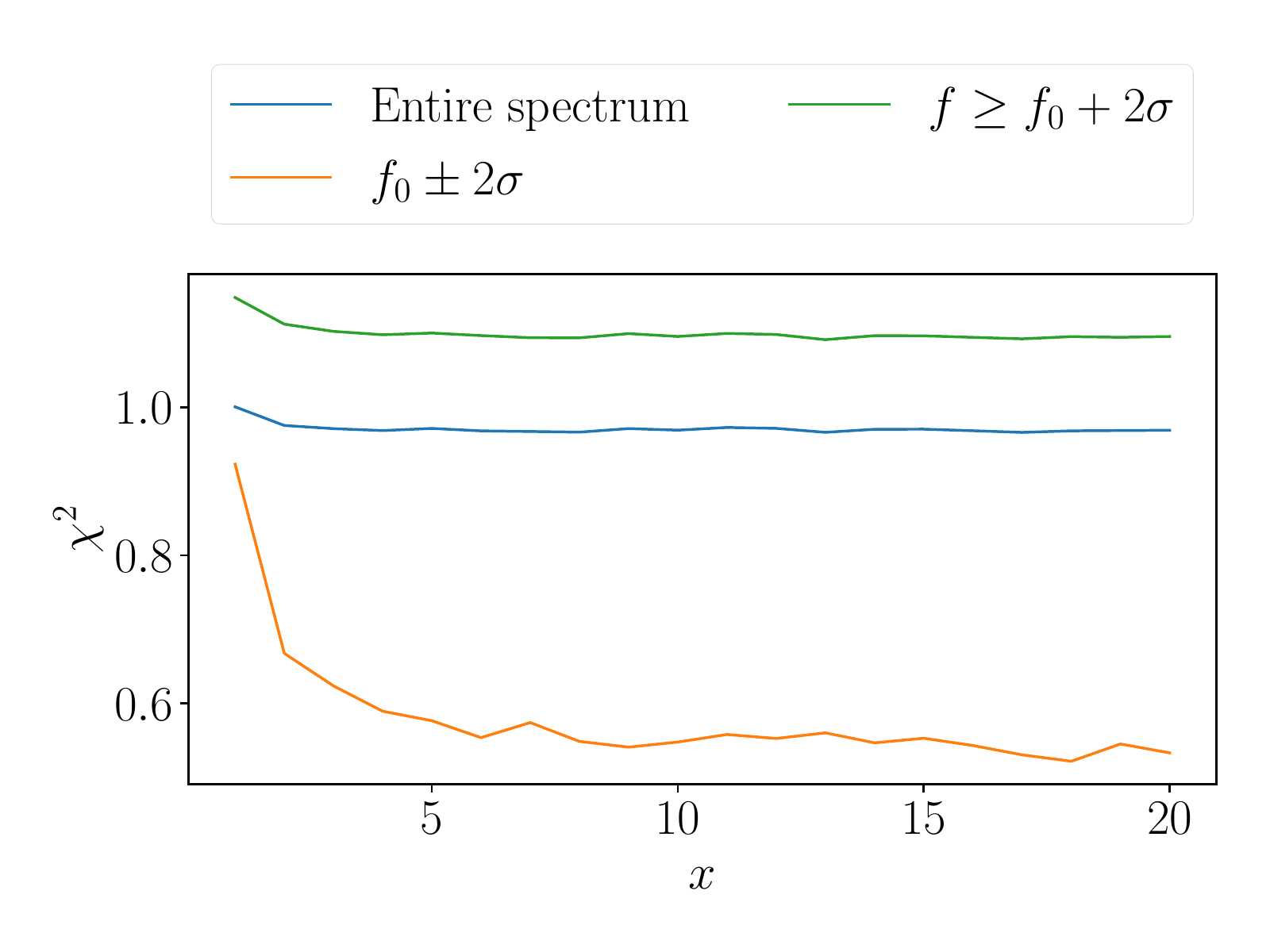}}
\subfigure{\includegraphics[width=\linewidth]{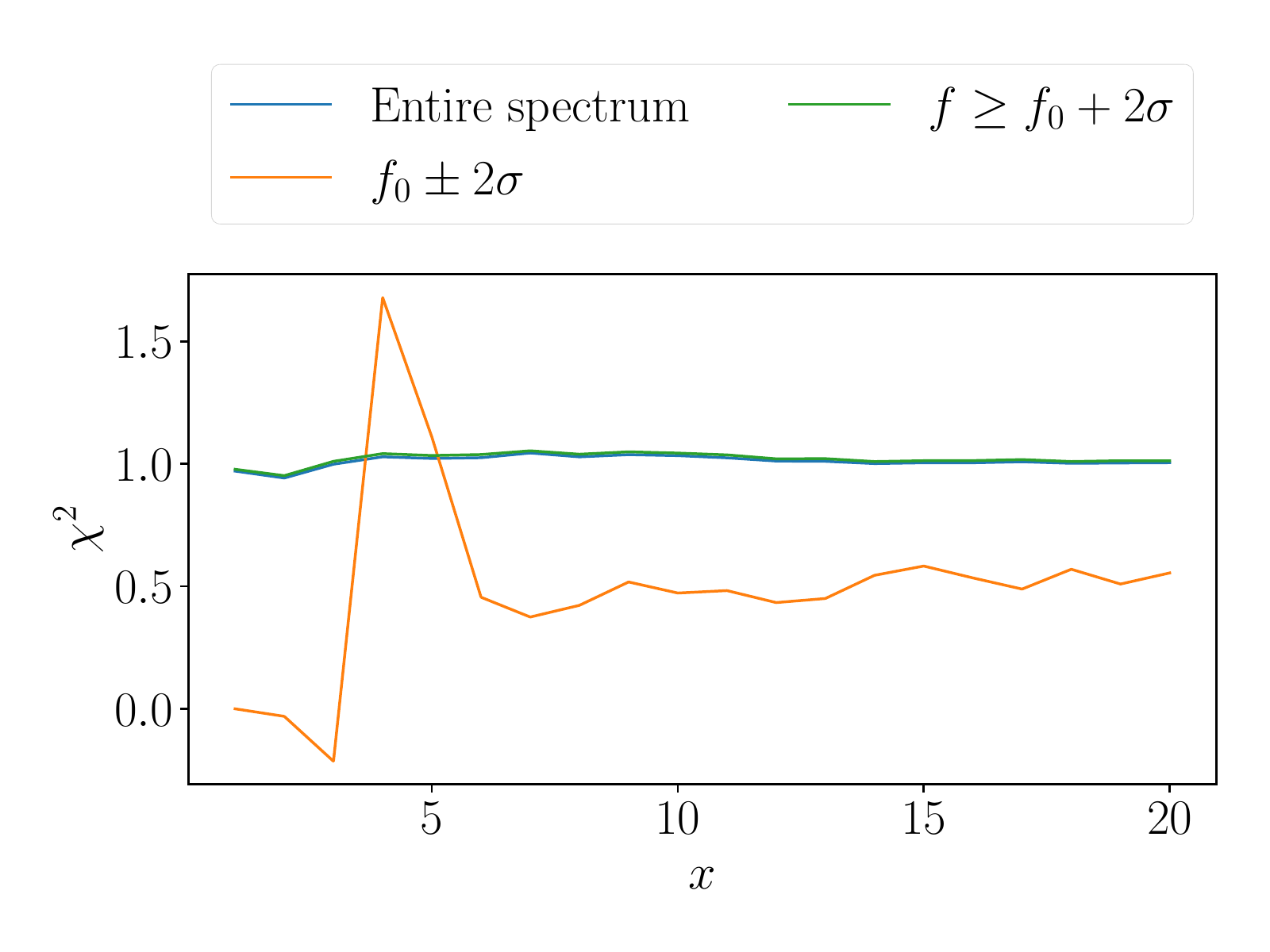}}
\caption{\citet{Nita2014} $\chi^2$-tests in the case of a QPO and white noise extended with zeros as described in Sec.~\ref{sec:c5:non_stat_qpo_in_wn} (top), and the case of QPO, red noise and white noise extended with white noise as described in Sec.~\ref{sec:c5:non_stat_qpo_in_non_stat_red_noise} (bottom).
We see in the top panel that the overall fit (blue) and the fit for frequencies above the QPO frequency (green) have $\chi^2 \approx 1$, indicating a good fit.
The local $\chi^2$-test around the QPO frequency $f_0$ indicates a good fit only for $x=1$ and falls much below that for $x > 1$, indicating overfitting.
This implies that the local $\chi^2$-test can detect non-stationarity bias in this instance.
In the bottom panel, we display some of the limitations of the local $\chi^2$-test when we deal with narrow QPOs.
For $x \lesssim 6$ the local $\chi^2$-test (orange) around $f_0$ swings widely as we have too few frequency bins within the range.
Above the $x \approx 6$ enough frequency bins accumulate within the range and we detect the overfitting.
We also note that the $\chi^2$-test for the entire PSD (blue) and frequencies above the QPO frequency (green) indicate an overall good fit despite us overfitting the QPO}
\label{fig:chi_square}
\end{figure}

One noticeable property of the bias we have found is that it passes regular tests employed to detect fitting validity.
~\citet{Nita2014} demonstrates a $\chi^2$-like test for exponentially distributed data
\begin{equation}\label{eq:chi_square}
    \chi^2_{\nu} = \frac{1}{\nu}\sum_{j=1}^{N/2}\left(1 - \frac{I_j}{\hat{S}_j}\right)^2 \, ,
\end{equation}
where $\nu$ are the number of degrees freedom and $\hat{S}$ is the maximum likelihood PSD. 
Analogously to the regular $\chi^2$-test, $\chi^2_{\nu}\approx 1$ indicates an appropriate fit to the data, whereas $\chi^2_{\nu} > 1$ indicates underfitting, and  $\chi^2_{\nu} < 1$ overfitting.
% Equation~\ref{eq:chi_square} is supposed to detect bad fits and return a value less or greater than one if we overfit or underfit the data, respectively.
Poor fits should in general occur with non-stationary behavior in the red noise or QPO dominated part of the PSD.
That is because neighboring bins will be covariant and a fit may either align more closely or further away than should be possible for independent bins, though $\chi^2 \approx 1$ may still randomly occur.
However, in practice, non-stationarities will affect the low-frequency part of the PSD while high frequencies are dominated by white noise.
Since frequency bins are linearly spaced, the vast majority of all bins are almost always fitted well in terms of the $\chi^2$-statistic if the periodogram reaches a white noise floor at high frequencies.
As we show in Fig.~\ref{fig:chi_square}, the $\chi^2$-value for the entire PSD is very close to the $\chi^2$-value for all the frequencies above the QPO frequency plus twice its width.
For comparison, setting $f_0 = \SI{1}{\hertz}$ with a \SI{40}{\hertz} sampling frequency, as it is the case in the bottom panel of Fig.~\ref{fig:chi_square}, about $95\%$ off all frequency bins are greater than $f_0$.
Thus even if we overfit locally around the QPO, it is unlikely that this greatly impacts the overall $\chi^2$-value.

We propose a local $\chi^2$-test, where we calculate the $\chi^2$-value for just the frequencies within $2\sigma$ around the QPO frequency.
We show this test in Fig.~\ref{fig:chi_square} for data sets used in Secs.~\ref{sec:c5:non_stat_qpo_in_wn} (top) and \ref{sec:c5:non_stat_qpo_in_non_stat_red_noise} (bottom).
We note in the top panel that we obtain a reasonable $\chi^2 \approx 0.92$ for the local test if we just look at the stationary time series at $x=1$, and obtain $\chi^2 < 0.6$ for large $x$. 
This method has the downside that it will not work if there are only very few bins available as is the case for $x \lesssim 6$ in the bottom panel of Fig.~\ref{fig:chi_square}.
We need at least seven data points to compensate for the six degrees of freedom in the $S^{\mathrm{RWQ}}$ model.
If more bins are available, this test can reliably flag whether the QPO has been overfitted or underfitted.
This test is required but not sufficient to show that non-stationarities are not impacting the analysis.
In general, we may find $\chi^2 \approx 1$ even for QPOs that are strongly affected by non-stationarity bias.
%Another issue of the local $\chi^2$-test is that it may flag the event not just because of non-stationarities, but also because the QPO model is unsuitable for the specific feature being investigated.
For example, a time series may in principle contain two QPOs at similar frequencies and non-stationarity bias.
In that instance, the overfitting due to covariance of neighboring bins may be compensated by underfitting of the underlying multiple QPO shapes and yield $\chi^2 \approx 1$ for the local test.
% Thus, interpreting $\chi^2 > 1$ as either being underfitted because of non-stationarities or because of underlying spectral features is not possible.

\section{Solar flare data}\label{sec:c5:solar_flare}
QPOs are a regularly reported phenomenon that have been observed in solar flares over a wide range of wavelengths for decades, from radio waves to EUV as well as soft and hard X-rays (see \citet{Nakariakov2009, VanDoorsselaere2016} for recent reviews).
In this domain, they are usually referred to as quasi-periodic pulsations (QPPs).
Due to their ubiquity, progress has been undertaken to automate their analysis using Whittle likelihoods, e.g. with the \textit{Automated Flare Inference of Oscillations} (AFINO) method~\citep{Inglis2015, Inglis2016, Hayes2020}.
An extensive catalog of AFINO analyses is available online\footnote{\url{https://aringlis.github.io/AFINO/}}.
AFINO defines models similar to $S^{\mathrm{RW}}$, $S^{\mathrm{RWQ}}$, and $S^{\mathrm{BW}}$, although it uses a Gaussian instead of a Lorentzian pulse as the QPO component, and a slightly different broken power-law model.
AFINO uses a computationally efficient \textsc{Scipy} fitting routine to determine the maximum likelihood values and calculate the $\Delta BIC$ with Eq.~\ref{eq:delta_bic}, though it does not calculate Bayes factors or posterior distributions.
The QPO model is ``strongly favored'' if $\Delta BIC < -10$, relative both to the red noise and broken power-law model.
This ensures a small number of false-positive values, as has been demonstrated on simulated data in~\citet{Broomhall2019}.
Additionally, AFINO performs the $\chi^2$-test for exponentially distributed data using Eq.~\ref{eq:chi_square}.
AFINO flags the results if the implied p-value from the $\chi^2_{\nu}$ test is below or above a set threshold.

\begin{figure}
    \centering
    \includegraphics[width=\linewidth]{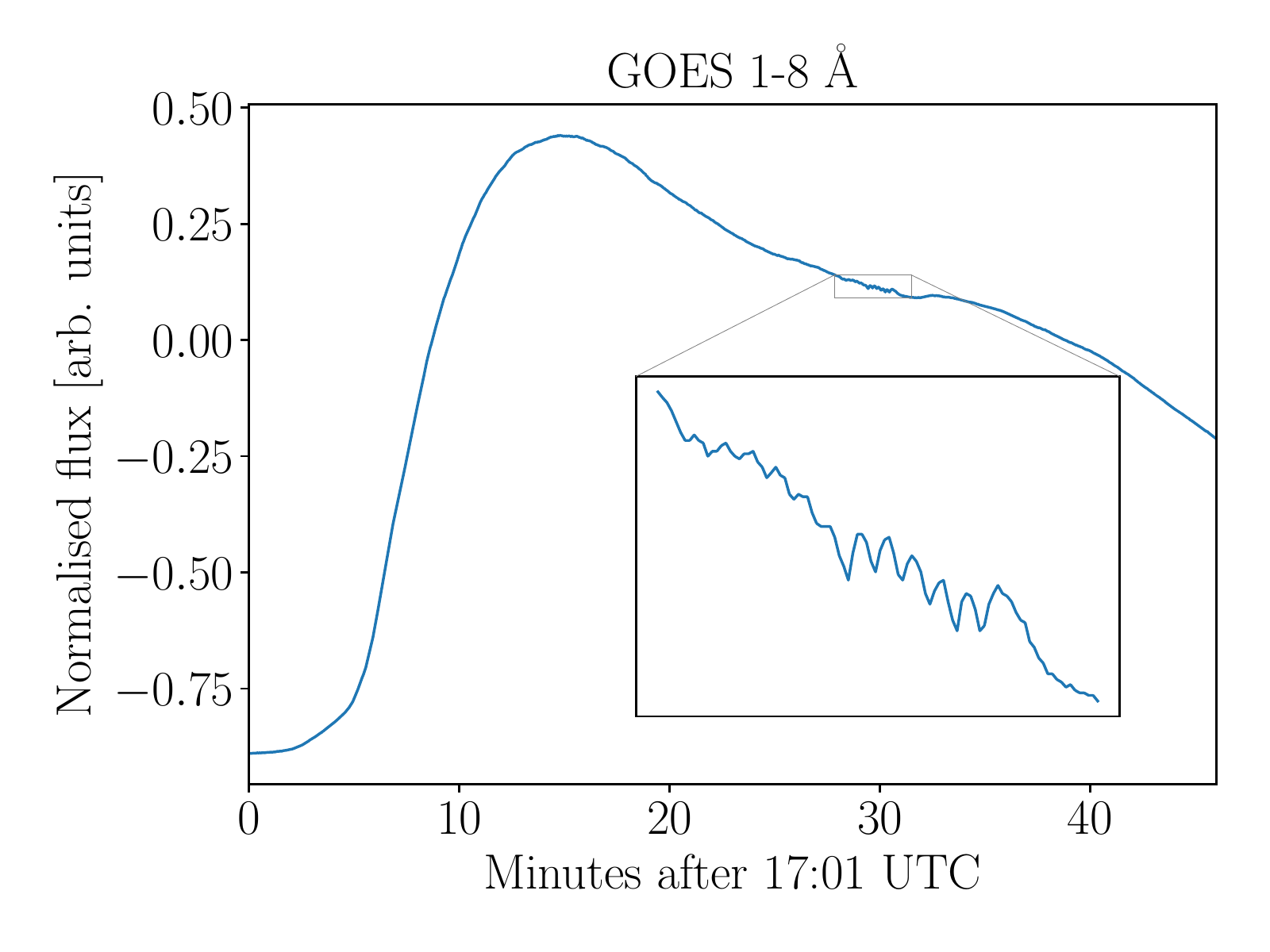}
    \caption{Solar flare x-ray lightcurve as observed with GOES in the \SIrange{1}{8}{\angstrom}  band. The figure shows the time selected as done with AFINO. The inset zooms in on the tail of the lightcurve where the QPO is clearly visible.}
    \label{fig:goes_time_series}
\end{figure}

For a comparative study on the impact of non-stationarity bias, it is instructive to consider some of the most significant QPOs.
This is because these QPOs are visually identifiable from the lightcurve, which guides us in their analysis.
We look at a solar flare observed by the X-Ray Sensor (XRS) onboard the GOES-15 satellite on May 12, 2013, at 20:17 UT for which AFINO reports the highest significance for a QPO with $\Delta BIC = -451.6$ relative to red noise and $\Delta BIC = -278.2$ relative to the broken power law.
This flare was of GOES magnitude M1.9 and originated from NOAA active region 11748, which was located on the East limb of the Sun at the time.
AFINO finds a high quality fit for the QPO model with $\chi^2 = 1.06$ but flags the fits with the broken power law and red noise due to their high $\chi^2$ values, likely due to the very pronounced QPO present at $P=\SI{12.6}{\second}$.
We show the 46 minutes long GOES x-ray lightcurve we analyze in Fig.~\ref{fig:goes_time_series}.
The QPO is visible by eye on the tail end of the distribution (inset) and persists for about \SI{200}{\second}, much shorter than the  \SI{2760}{\second} data segment.
This particular QPO may not be of solar origin, since it is not observed in other flare-observing instruments. However, a similar signature is coincidentally seen in the GOES magnetometer data, suggesting a possible artifact. Regardless of the origin, the QPO is clearly present and confined to a small portion of the data, providing an ideal test case.

\begin{figure}
    \centering
    \includegraphics[width=\linewidth]{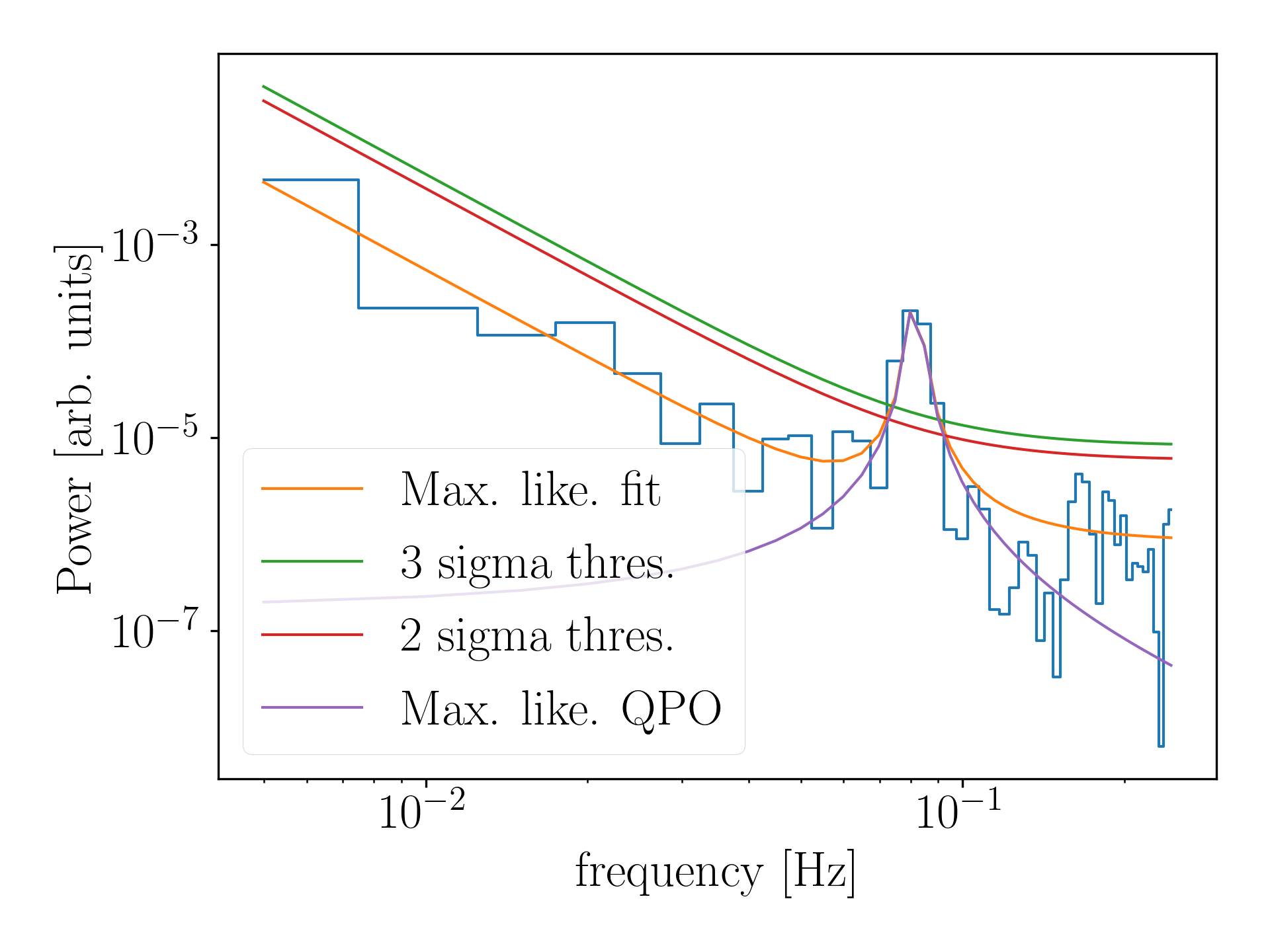}
    \caption{Maximum likelihood fit of the periodogram just for the second segment of the GOES lightcurve.
    The QPO is clearly visible and highly significant.
    We draw the $2\sigma$ and $3\sigma$ detection limits based on the Bonferroni-corrected frequentist statistics.
    We note that the QPO is located in the red noise dominated part of the PSD.}
    \label{fig:solar_flare_segment_2_max_like_fit}
\end{figure}

To demonstrate the presence of an artificial amplification of the SNR, we split the lightcurve into three segments.
The first segment covers the time before the QPO (\SIrange{0}{1680}{\second}) (after 20:17 UT), the second selects just the QPO (\SIrange{1680}{1880}{\second}), and the third segment selects everything after the QPO (\SIrange{1880}{2760}{\second}).
We perform Bayesian inference separately on each of these segments and the combined segment independently with the model $S^{\mathrm{RW}}$ and $S^{\mathrm{RWQ}}$.
We use the same priors as in the studies on simulated data which we list in Tab.~\ref{tab:priors}.
If the lightcurve were stationary, it would be valid to combine the Bayes factors and $f_0$ posteriors as they represent independent draws from the same distribution.
However, as we show in Tab.~\ref{tab:solar_flare_results}, we are unable to detect the reported QPO in the first and third segments.
There is some weak evidence towards a QPO in the first segment, though not at the reported \SI{12.6}{\second} period.
The third segment shows some weak support for a QPO with $\ln BF = 1.4$, but the maximum likelihood fit indicates that this is rather due to a broad feature in the periodogram that is better fitted with a broken power law.
As we show in Tab.~\ref{tab:solar_flare_results}, both the first and third segments are better fitted with $S^{\mathrm{BW}}$ than with $S^{\mathrm{RWQ}}$, which indicates that it is likely no QPO present.
In the second segment we find the QPO independently with a very high significance $\ln BF = 27.5$, and it is clearly visible in the fitted periodogram in Fig.~\ref{fig:solar_flare_segment_2_max_like_fit}.
Finally, we analyze the combined segments together and find the QPO with $\ln BF = 229.4$, a significance much higher than in segment two.
We note that we find the QPO at a slightly different period of $\SI{12.34}{\second}\pm\SI{0.06}{\second}$ compared to the value reported of AFINO.
This may be due to slightly different QPO modeling choices between AFINO and our method.

We perform the local $\chi^2$-test we introduced in Sec.~\ref{sec:c5:chi_square} for the entire lightcurve and find $\chi^2 = 0.29$ for the frequency bins surrounding the QPO.
This indicates that the QPO has been overfitted and is non-stationary.

\begin{table*}[t]
    \centering
    \begin{tabular}{c|c|c|c|c}
        Segment                       & $\ln BF$ $S^{\mathrm{RWQ}}$ & $\ln BF$ $S^{\mathrm{BW}}$ & $\Delta BIC$ $S^{\mathrm{RWQ}}$ & $\Delta BIC$ $S^{\mathrm{BW}}$ \\ \hline
        \SIrange{0}{1680}{\second} & $3.0 \pm 0.3$      & $6.8\pm 0.3$             &	-11.3						& -5.0 \\
        \SIrange{1680}{1880}{\second} & $27.5 \pm 0.2$     & $-0.7 \pm 0.2$           &	-66.8						& 7.8 \\
        \SIrange{1880}{2760}{\second} & $1.4 \pm 0.3$      & $7.7\pm0.2$              &	-0.7						& -6.8 \\
        \SIrange{0}{2760}{\second} & $229.4 \pm 0.3$    & $93.0 \pm 0.3$           &	-465.6						& -175.8 
        \end{tabular}
    \caption{Results from analyzing the selected segments of the solar flare detected by GOES. All errors are given based on $1-\sigma$ confidence or credible intervals.
    The segments are given in seconds after 20:17 UT.
    All $\ln BF$ and $\Delta BIC$ values are calculated relative to $S^{\mathrm{RW}}$.
    We reiterate that either a positive $\ln BF$ or a negative $\Delta BIC$ indicate that $S^{\mathrm{RWQ}}$ or $S^{\mathrm{BW}}$ are preferred over $S^{\mathrm{RW}}$. 
    We find broadly that the $\Delta BIC$ values in the last row are in agreement with what AFINO reported.
    Deviations are likely due to differences, in our model and slightly different data selection.
    While in all instances Bayes factors and $\Delta BIC$s give the same indication about the preferred model, the significance differs to some extent. 
    This may be in part due to our wide prior choices. }
    \label{tab:solar_flare_results}
\end{table*}

It is evident that the solar flare segment we have analyzed contains a QPO and an overestimation of its significance is not as critical as it would be for a marginal detection.
Trying to find an instance where a marginal detection was turned into a very confident detection due to non-stationarity bias would be much harder.
We would not be able to determine the location of the QPO in the lightcurve by eye.
Instead, we would have to take a systematic approach and split the lightcurve into several smaller segments and determine if the QPO exists in them individually.
Of course, the significance in the individual segments would always be lower, so it would be hard to determine whether this is due to the reduced non-stationarity bias or because we split a persistent QPO into multiple segments.
As mentioned previously, we do not expect AFINO to report many false detections, as has been established in ~\citet{Broomhall2019}.
This is due to its conservative detection threshold $\Delta BIC < -10$.
Thus, if AFINO is coupled with a more careful data selection method it could be possible to find more QPOs.

The non-stationarity bias in solar flare data arises most likely as a combination of the effects we describe in Sec.~\ref{sec:c5:pop}, i.e. a non-stationary QPO, non-stationary noise processes, and a deterministic overall flare shape all contribute to some extent.
We can not easily discern what the impact of each of these effects is.

Non-automated analyses of QPOs in solar flares are also likely less affected than the stretch of GOES data we analyze in this section.
Typically, parts of the lightcurve that are of interest are selected manually for detailed analyses \citep[e.g.~][]{Hayes2019}, though this by no means provides a guarantee that such results are without bias.
As we have established previously, the $\ln BF$ or the $\Delta BIC$ can grow linearly with $x$, thus even a mild overselection of the lightcurve can lead to erroneous detections and inferences.
\section{Discussion and outlook}\label{sec:c5:discussion}
In this paper, we show that analyses based on Whittle likelihoods likely overstates the significance of QPOs in lightcurves if the QPO or noise features in time series are non-stationary.
Specifically, this is the case if data is included in the time series that contains only white noise when the QPO is in the red-noise-dominated part of the PSD.
The effect can also occur if the QPO is in the white-noise-dominated part of the PSD, but the white noise is non-stationary (e.g. if the noise arises due to photon counting) and we include weak white noise from parts of the lightcurve that have low photon counting rates.
Selecting an appropriate time segment for the analysis of transients is thus important but remains difficult to do generically and without bias.
A scientist's natural intuition may be to select data conservatively, i.e. select a longer time series while being aware that the QPO may only be present for a part of it.
The erroneous reasoning may be that by selecting a longer segment they are not ``cherry-picking'' the segment with the most pronounced QPO features.
As we demonstrate in this paper, this choice may increase the significance of the QPO artificially and lead to false detections due to the non-stationarity bias.

There are some possible mitigation strategies that one may apply when the use of periodograms is still preferred.
However, this means that it is a lot harder to devise a generic process with which to analyze time series.
Firstly, if we suspect that the QPO is indeed only present for part of the transient, and the white noise level is relatively low compared to the low-frequency continuum, as is indeed the case for x-ray lightcurves of solar flares, it is reasonable to seek out the part of the lightcurve that looks by eye most likely to contain a QPO, and then set the limits on the start and end of the segment manually.
However, this method requires significant human supervision and thus will not scale to large data sets.
It is open to other forms of selection bias, too.
Secondly, for such transients, we also suggest splitting the lightcurve up into several parts to see if one can detect the QPO in all segments or only in some of them.
For the final analysis, one should only use segments if there is some evidence for the QPO.
Next, for shorter transients where we suspect the QPO to be present for most of the time, we should conservatively trim the lightcurve after the onset and before the end of the segment.
This way we are less likely to make the lightcurve non-stationary.
Finally, we have found that regular $\chi^2$-like tests are unsuitable to detect non-stationary bias since only a few bins around the QPO are affected.
The overall $\chi^2$-value is mostly determined by the far greater share of frequency bins that lie in the white-noise-dominated part of the PSD.
We outline that alternatively calculating $\chi^2$ based solely on the frequencies surrounding the QPO can detect overfitting, which is a hallmark of statistically non-independent frequency bins.

Aside from periodograms, other methods that can be used for the search for QPOs.
Wavelet transforms are a popular way to analyze the time series by convolving a wavelet function with a time series of interest~\citep{Torrence1998}.
Wavelet analyses are not restricted to stationary data sets and may help us to find the specific times when a QPO occurs.
However, the statistics of detecting QPOs with wavelets in the presence of red noise remains contentious since detrending methods are likely to lead to false detections~\citep{Auchere2016}.
Alternatively, we can avoid frequency-domain methods altogether and model the time series as a Gaussian process with some mean function.
The main drawback of Gaussian processes is that they have computational complexity $\mathcal{O}(N^3)$ and thus are only suitable for short time series in the general case~\citep{Rasmussen2006}, though progress has been made to reduce complexity to $\mathcal{O}(N)$ for stationary, complex exponential kernels or combinations thereof~\citep{Foreman-Mackey2017}.
We note that we expect the same non-stationarity bias if we apply a stationary Gaussian process kernel to a non-stationary time series since the Whittle likelihood explicitly derives from a Gaussian process likelihood.
We have also found this result empirically in some preliminary analyses (Huebner et al. 2021, in prep).

One advantage of Gaussian processes is that they allow us to fit the overall shape of the transient and the stochastic process simultaneously, instead of lumping the shape of the transient into the red noise, which can help us to prevent the bias demonstrated in Sec.~\ref{sec:c5:non_stat_qpo_in_transient}.
It thereby allows us to avoid nonparametric methods that are likely to create false periodicities~\citep{Auchere2016}.
Additionally, the Gaussian process likelihood can explicitly take in the known variance for flux values within a lightcurve if they are available.
Thus, they also provide a natural resolution to the bias shown in App.~\ref{sec:c5:det_process_poiss}.
Finally, Gaussian processes allow us to model relatively simple, non-stationary extensions to the fast, stationary models that are popular now (Huebner et al. 2021, in prep).
Further development of Gaussian process methods may eventually allow us to create more complex models of QPOs and their possible intermittency within transients.
\paragraph{Acknowledgments}
This work is supported through ARC Centre of Excellence CE170100004.
P.D.L. is supported through the Australian Research Council (ARC) Future Fellowships FT160100112, ARC Discovery Project DP180103155.
D.H. is supported by the Women In Science Excel (WISE) programme of the Netherlands Organisation for Scientific Research (NWO).
This work was performed on the OzSTAR national facility at Swinburne University of Technology. 
The OzSTAR program receives funding in part from the Astronomy National Collaborative Research Infrastructure Strategy (NCRIS) allocation provided by the Australian Government.

\appendix

\section{Priors and simulated data parameter tables}\label{sec:c5:injection_param_table}
We list the priors and parameters used for all the studies in Secs.~\ref{sec:c5:injection_study} and~\ref{sec:c5:solar_flare} in Tabs.~\ref{tab:priors} and ~\ref{tab:injection_params}.
\begin{table*}[h]
    \centering
    \begin{tabular}{c|c|c|c|c}
         Parameter & Description & Prior & Minimum & Maximum \\  \hline
         $A$ & Red noise amplitude & \texttt{LogUniform} & $\exp(-30)$ & $\exp(30)$\\ 
         $\alpha$, $\alpha_{1}$, $\alpha_{2}$ & red noise / BPL spectral index & \texttt{Uniform} & $0$ & $10$\\
         $B$ & QPO amplitude & \texttt{LogUniform} & $\exp(-60)$ & $\exp(60)$\\
         $f_0$ & QPO central freq. & \texttt{LogUniform} & $2\Delta f $ & $ f_{\max}$\\
         $\sigma$ & QPO HWHM & \texttt{ConditionalLogUniform} & $\Delta f/\pi $ & $ 0.25f_0$\\
         $C$ & White noise amplitude & \texttt{LogUniform} & $\exp(-30)$ & $\exp(30)$\\
         $\delta$ & Breaking freq. for BPL & \texttt{LogUniform} & $\Delta f$ & $f_{\max}$
    \end{tabular}
    \caption{Parameters and priors used throughout all studies in this paper.
    Priors are referred to by their implementation in \bilby.
    We select wide priors in $A$, $B$, and $C$ for simplicity. Although the selection of priors this wide may have a slight impact on the Bayes factor calculation, it will not qualitatively change the scaling of Bayes factor with $x$. 
    We also set a wide prior on $\alpha$, or $\alpha_{1}$ and $\alpha_{2}$, and enforce $\alpha_2 < \alpha_1$ using a \texttt{Constraint} prior in \bilby to avoid degeneracies.
    The prior on $f_0$ is motivated by the available frequencies in the periodogram.
    We set the minimum $f_0$ at twice the difference between neighboring frequencies $\Delta f$ to ensure better convergence, and refer to the highest frequency in the periodogram as $f_{\max}$.
    We set $\sigma$, the width of the QPO at half width half maximum (HWHM), to be conditional uniform in log between $\Delta f/pi$, i.e. on the scale of a single frequency bin, and $0.25 f_0$, to prevent it from converging towards wider features in the periodogram.
    For the broken power law analysis of solar flare data, we use priors for $\alpha_{1,2}$ identical to $\alpha$, and the listed $\delta$ prior}\label{tab:priors}
\end{table*}

\begin{table*}[h]
    \centering
    \begin{tabular}{c||c|c|c|c|c}
         Parameter &  Sec.~\ref{sec:c5:non_stat_qpo_in_wn} zeros & Sec.~\ref{sec:c5:non_stat_qpo_in_wn} white noise & Sec.~\ref{sec:c5:non_stat_qpo_in_transient} & Sec.~\ref{sec:c5:non_stat_qpo_in_non_stat_red_noise} & App.~\ref{sec:c5:det_process_poiss}\\  \hline
         $A$ & - & - & - & 4 & -\\ 
         $\alpha$ & - & - & - & 2 & -\\
         $B$ & 3  & 15 & 80 & 8 & 100\\
         $f_0$ & 5 & 5 & 1 & 1 & 5\\
         $\sigma$ & \SI{0.1}{\hertz} & \SI{0.1}{\hertz} & \SI{0.01}{\hertz} & \SI{0.02}{\hertz} & \SI{0.1}{\hertz} \\
         $C$ & 2 & 2 & 2 & 2 & -\\
         $A_{\mathrm{flare}}$ & - & - & 20000 & - & \SI{1000}{\per\second}\\
         $t_0$ & - & - & \SI{70}{\second} & - & \SI{200}{\second}\\
         $\tau_\mathrm{r}$ & - & - & \SI{10}{\second} & - & - \\
         $\tau_\mathrm{f}$ & - & - & \SI{20}{\second} & - & - \\
         $\sigma_{\mathrm{flare}}$ & - & - & - & - & \SI{20}{\second} \\
         $c_0$ & - & - & - & - & \SI{10}{\per\second} \\
         Segment length & \SI{400}{\second} & \SI{400}{\second} & \SI{200}{\second} & \SI{400}{\second} & \SI{400}{\second} \\
         Sampling frequency & \SI{40}{\hertz} & \SI{40}{\hertz} & \SI{40}{\hertz} & \SI{40}{\hertz} & \SI{40}{\hertz}
    \end{tabular}
    \caption{Values used to create the simulated data in Sec.~\ref{sec:c5:injection_study}.
    The column heads refer to the subsection in which this set of parameters was used.
    We use two different sets of parameters in Sec.~\ref{sec:c5:non_stat_qpo_in_wn} depending on whether we extend using zeros or more white noise.
    }\label{tab:injection_params}
\end{table*}

\section{Deterministic processes impact white noise}\label{sec:c5:det_process_poiss}
White noise observed in astrophysical lightcurves using photon counting does not arise due to intrinsic properties of the source, but rather due to the Poissonian noise nature of photon counting.
Concretely, given a rate $\lambda$, the distribution of the number of observed photons $k$ in a unit of time is Poissonian
\begin{equation}
    \mathrm{Pois}(k; \lambda) = \frac{\lambda^k e^{-\lambda}}{k!} \, .
\end{equation}
Photon counting noise thus scales proportionally to the standard deviation of the Poisson distribution $\sqrt{\lambda}$.
This relation implies that photon-counting noise throughout a transient does not remain constant.
We demonstrate a limit of this in Sec.~\ref{sec:qpos_in_stationary_noise} when we consider what happens if we extend a time series with zeros.
This limit corresponds to the case in which a detector sees no photons at all.
In realistic detectors, we will not reach this limit as there are at least some background photons, though we can get close enough to it for it to lead to wrong inferences about QPO significance.

The above effect points to a deficiency in periodograms more generally.
For the calculation of a periodogram we use the values $x(t_i)$ of a discrete time series, but we do not include their variances $\Delta x(t_i)$ if they are available.
In effect, this implies a loss of information as the white noise is already encoded in the variance associated with the photon counts.
Instead, we infer it after the fact via power spectral density estimation independently and lose knowledge about non-stationarity in the white noise.

% Though this effect is present in any astrophysical setting that involves photon counting, we separate the treatment into an example to demonstrate that this effect is different from the ones involving low frequency noise components.

\begin{figure}
\centering
\subfigure{\includegraphics[width=0.8\linewidth]{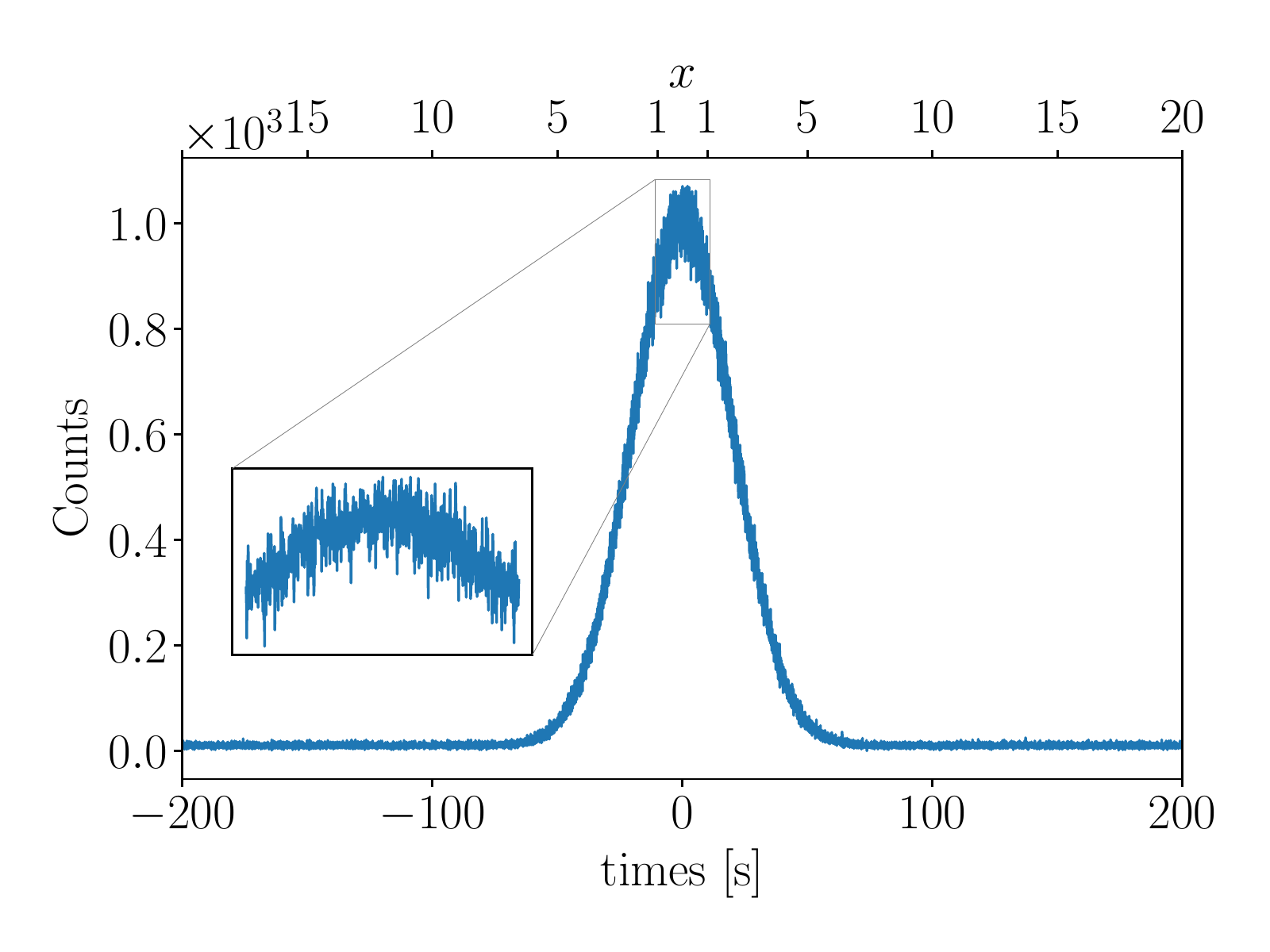}}
\subfigure{\includegraphics[width=0.8\linewidth]{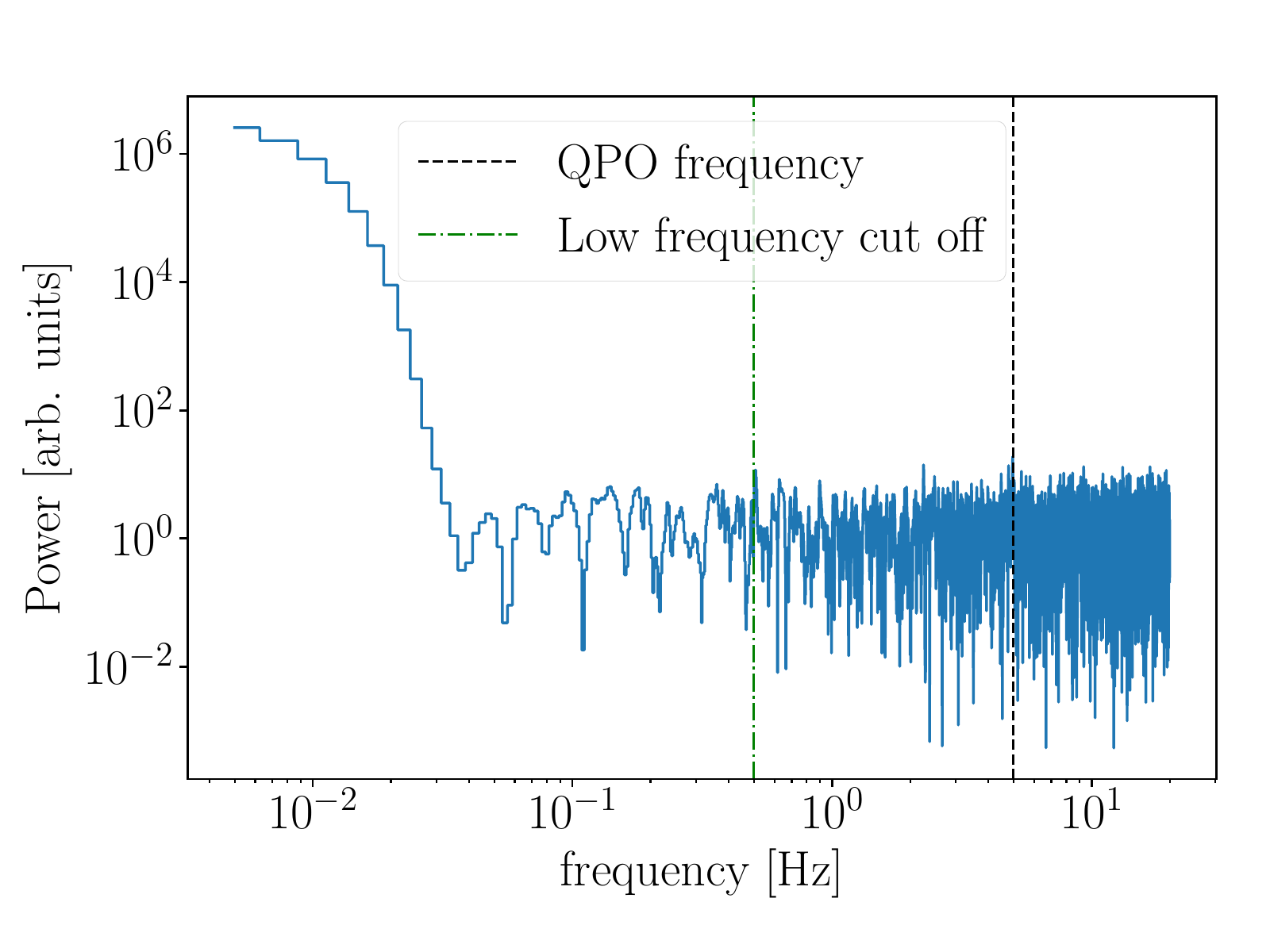}}

\caption{Simulated data of a non-stationary QPO in a deterministic transient flare shape in the presence of white photon counting noise. 
The time series (top) shows the dominant transient and the QPO that is present for \SI{20}{\second} on the top of the flare (inset).
The periodogram for the entire \SI{400}{\second} time series (bottom) has a relatively wide ($\sigma=0.1$) QPO at \SI{5}{\hertz} (barely visible).
For our analysis, we cut off all frequencies below \SI{0.5}{\hertz} to avoid any effects arising from the power in low frequencies due to the flare shape.
}
\label{fig:inj_qpo_in_poiss_data}
\end{figure}

To demonstrate this effect, we construct a transient similar to Sec.~\ref{sec:c5:non_stat_qpo_in_transient}, though in a way that its low-frequency contributions do not overlap with the QPO frequency, as to demonstrate that this effect arises due to non-stationarity in the white noise.
Concretely, we construct the transient by generating a Gaussian profile plus a constant
\begin{equation}
    f(t; A_{\mathrm{flare}}, t_0, \sigma_{\mathrm{flare}}) = A_{\mathrm{flare}} \exp \left(-\frac{(t-t_0)^2}{2\sigma_{\mathrm{flare}}^2} \right) + c_0 \, ,
\end{equation}
where $A_{\mathrm{flare}}$ is the amplitude, $t_0$ is the time of the peak, $\sigma_{\mathrm{flare}}$ is the width, and the constant $c_0$ represents a possible background count rate.
We add a \SI{20}{\second} non-stationary QPO in the same manner as we describe in Sec.~\ref{sec:c5:non_stat_qpo_in_transient}.
We create photon counts, which we show in Fig.~\ref{fig:inj_qpo_in_poiss_data}, by simulating the Poisson process using the implementation in \texttt{scipy.stats.poisson.rvs}.
The specific parameters of this time series are listed in Tab.~\ref{tab:injection_params}.
By construction, the QPO is set on the top of the Gaussian profile, which corresponds to the highest level of photon counting noise.
The Gaussian profile, unlike the exponential from Sec.~\ref{sec:c5:non_stat_qpo_in_transient}, contributes powers at lower frequencies than the exponential, roughly up to $1/\sigma_{\mathrm{flare}} \approx \SI{0.05}{\hertz}$.
Additionally, we set our QPO frequency at \SI{5}{\hertz} and cut off frequencies below \SI{0.5}{\hertz} so that we can avoid using a red noise component in our modeling in the same way we did in Sec.~\ref{sec:c5:non_stat_qpo_in_wn}.

\begin{figure}
    \centering
    \includegraphics[width=0.8\linewidth]{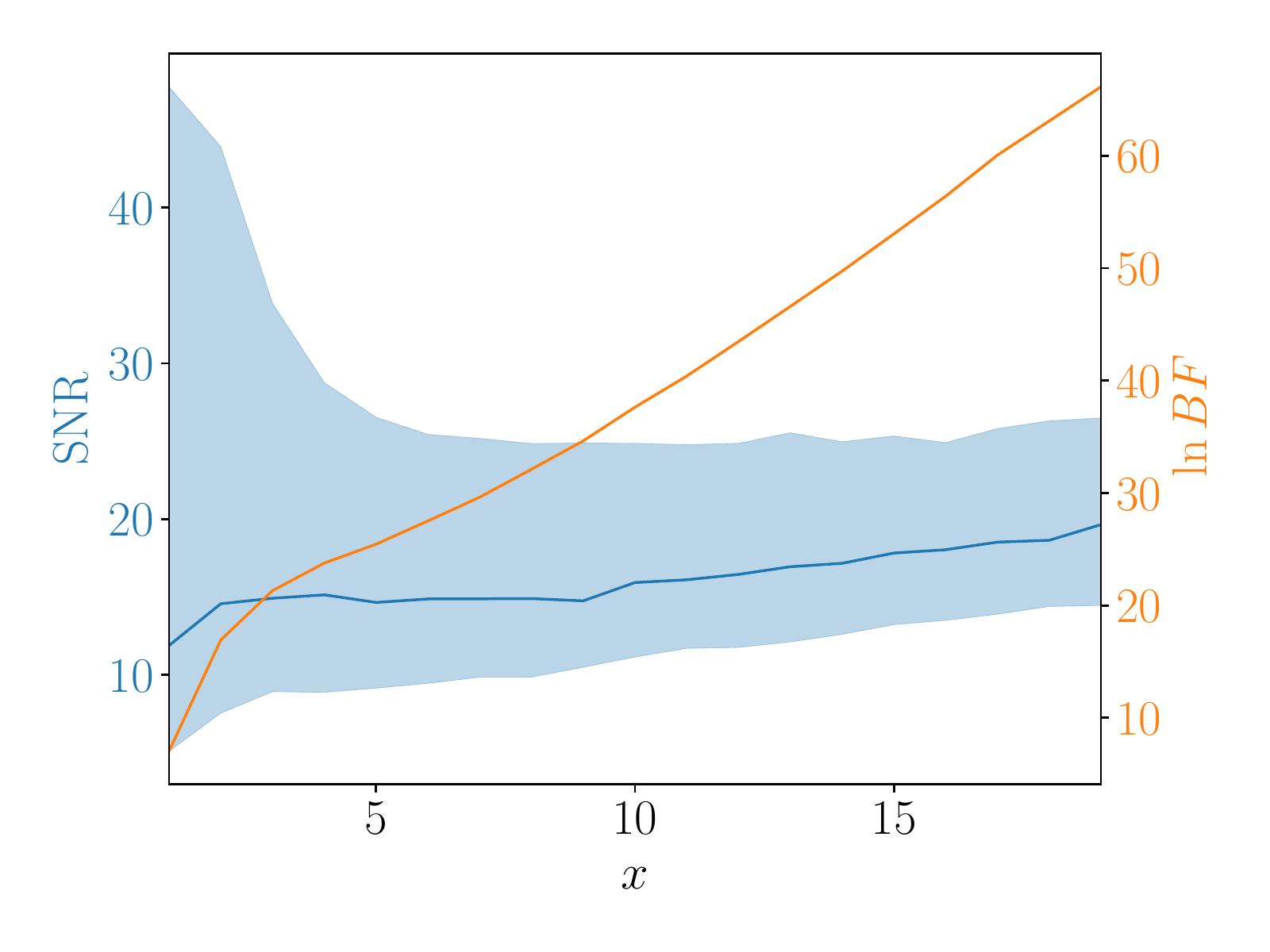}
    \caption{We calculate the Bayes factors for the presence of a QPO in a transient flare for different extension factors with $x=1$ corresponding the \SI{20}{\second} inset from the top of Fig.~\ref{fig:inj_qpo_in_poiss_data}.
    While we already start with very strong evidence of a QPO at $x=1$, the Bayes factors continue to increase as we extend the time series to include more of the transient. 
}
    \label{fig:inj_qpo_in_poiss_ln_bfs}
\end{figure}

On the tails of the flare, we expect the same effect as we observe when we extend with zeros in our other simulations, though Fig.~\ref{fig:inj_qpo_in_poiss_ln_bfs} clearly shows that we artificially increase the significance already with very small extension factors.
This is possible since the flare quickly falls off on either side and the added photon counting noise does not compensate for the increased number of frequency bins that contain the QPO.
Overall, this effect would be hard to account for just based on better methods for setting data cuts because varying levels of white noise are expected in many transient lightcurves.

\bibliography{QPOs}
\bibliographystyle{aasjournal}
\end{document}